\documentclass[a4paper,11pt]{article}
\usepackage{pos}
\usepackage{enumitem}
\usepackage[T1]{fontenc}
\usepackage{textcomp}
\usepackage{graphicx}
\usepackage{hyperref}
\newcommand{\orcid}[1]{\unskip\protect\href{https://orcid.org/#1}{\protect\includegraphics[width=8pt,clip]{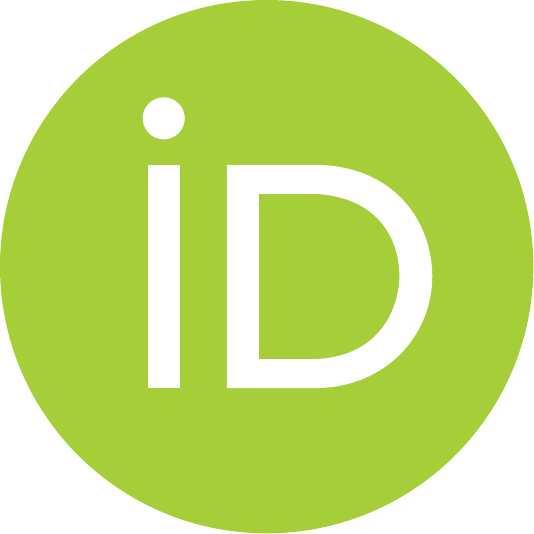}}}

\title{Dark matter searches in dwarf spheroidal galaxies with the Cherenkov Telescope Array}
 \ShortTitle{DM searches in dSphs with CTA}

\author*[a,b]{F. G. Saturni}
\author[c,d]{M. Doro}
\author[e,f]{A. Morselli}
\author[e,f]{G. Rodr{\'i}guez-Fern{\'a}ndez}
\author[g]{\\for the CTA Consortium}

\affiliation[a]{INAF -- Osservatorio Astronomico di Roma, Via Frascati 33, I-00078 Monte Porzio Catone (RM), Italy}

\affiliation[b]{ASI -- Space Science Data Center, Via del Politecnico s.n.c., I-00133 Roma, Italy}

\affiliation[c]{Universit{\`a} degli Studi di Padova -- Dip. di Fisica ed Astronomia ``G. Galilei'', Via Marzolo 8, I-35131 Padova, Italy}

\affiliation[d]{INFN -- Sezione di Padova, Via Marzolo 8, I-35131 Padova, Italy}

\affiliation[e]{Universit{\`a} di Roma ``Tor Vergata'' -- Dip. di Fisica, Via della Ricerca Scientifica 1, I-00133 Roma, Italy}

\affiliation[f]{INFN -- Sezione di Roma Tor Vergata, Via della Ricerca Scientifica 1, I-00133 Roma, Italy}

\affiliation[g]{\url{www.cta-observatory.org/consortium\_authors/}}

\emailAdd{francesco.saturni@inaf.it}

\abstract{
Dark matter (DM) is one of the major components in the Universe. However, at present its existence is still only inferred through indirect astronomical observations. DM particles can annihilate or decay, producing final-state Standard Model pairs that subsequently annihilate into high-energy $\gamma$-rays. The dwarf spheroidal galaxies (dSphs) in the Milky Way DM halo have long been considered optimal targets to search for annihilating DM signatures in GeV-to-TeV $\gamma$-ray spectra due to their high DM densities (hence high astrophysical factors), as well as the expected absence of intrinsic $\gamma$-ray emission of astrophysical origin. For such targets, it is important to compute the amount of DM in their halos in a consistent way to optimize the $\gamma$-ray data analysis. Such estimates directly affect the observability of DM signals in dSphs, as well as the DM constraints that can be derived in case of null detection. In this contribution, we present the results on the sensitivity of the Cherenkov Telescope Array (CTA) for DM annihilation and decay searches using planned observations of the Milky Way dSphs. We select the most promising targets among all presently known dwarf satellites, providing new determinations of their expected DM signal. This study shows an improvement of approximately an order of magnitude in sensitivity compared to current searches in similar targets. We also discuss the results in terms of cuspy and cored DM models, and investigate the sensitivity obtained by the combination of observations from different dSphs. Finally, we explore the optimal strategies for CTA observations of dSphs.
}

\ConferenceLogo{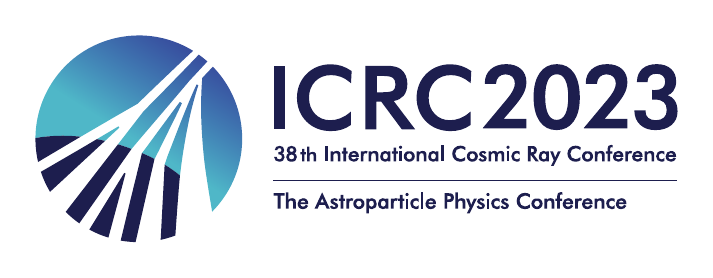}

\FullConference{%
The 38th International Cosmic Ray Conference (ICRC2023)\\
  26 July -- 3 August, 2023\\
  Nagoya, Japan}

\begin{document}
\maketitle

\section{Introduction}
\label{sec:intro}

The problem of establishing the nature of dark matter (DM; \cite{zwi33}) is one of the major open challenges in modern astrophysics. Several efforts on the side of elementary particles -- e.g., weakly interacting massive particles, (WIMPs; e.g., \cite{apr18}) or axion-like particles (ALPs; e.g., \cite{bat21}) -- have been made to identify plausible DM candidates. However, the parameter space covered by such candidates ranges over several orders of magnitude in masses and cross sections (e.g., \cite{ros04}). The current framework for the astronomical searches for DM signals is based on the possibility that DM particles annihilate or decay \cite{ber98} into Standard Model (SM) pairs that subsequently produce final-state $\gamma$-ray photons \cite{cem11,cir11}.

The concordance cosmological model predicts that the formation of visible astrophysical structures has been guided by the gravitational potential of previously formed DM overdensities. In particular, the dwarf spheroidal galaxies (dSphs; \cite{may10}) are highly DM-dominated and relatively nearby environments with respect to other cosmological DM reservoirs \cite{eva04,str08}. Therefore, they configure as one of the primary targets for observations aimed at detecting potential observable signals from particle DM. Nearby dSphs have already been the subject of extensive studies with currently operating imaging atmospheric Cherenkov telescopes (IACTs; \cite{ver18,mag22}), and are also optimal targets for next-generation IACTs such as the Cher\-enkov Telescope Array (CTA; \cite{dor13,cta19}). In this contribution, we present the most updated CTA sensitivities to DM searches in dSph halos, based on novel derivations of their expected DM content. The dSph astrophysical factors for DM annihilation $J_{\rm ann}$ and decay $J_{\rm dec}$ \cite{eva04} estimated in this way are in turn used to compute the expected DM $\gamma$-ray signal intensities and to rank such targets in view of their observation.

\section{Astrophysical factors for dwarf spheroidal galaxy halos}
\label{sec:jfac}
The $\gamma$-ray flux produced by annihilating or decaying DM can be written as:
\begin{equation}\label{eqn:dm_flux}
\frac{d\Phi_\gamma}{dE_\gamma\,d\Omega}=
\begin{cases}
 \dfrac{\langle \sigma v \rangle}{8 \pi m_{\rm DM}^2} \; \sum_i {\rm BR}_i \dfrac{dN^{(i)}_\gamma}{dE_\gamma} \;\cdot\; \dfrac{dJ_{\rm ann}}{d\Omega} %\qquad \rm{Annihilating~DM} \\
\\
\dfrac{1}{4 \pi m_{\rm DM}} \; \sum_i \dfrac{1}{\tau_i} \dfrac{dN^{(i)}_\gamma}{dE_\gamma}\;\cdot\; \dfrac{dJ_{\rm dec}}{d\Omega} %\qquad \rm{Decaying~DM} 
\end{cases}
\end{equation}
where $\langle \sigma v \rangle$ is the thermally-averaged DM annihilation cross section in case of annihilating DM and $\tau_i$ is the particle lifetime for DM decaying into the $i$-th SM channel, $m_{\rm DM}$ is the DM particle mass, $dN_\gamma/dE_\gamma$ is the number of photons produced during one interaction at a given energy $E_\gamma$ with a given branching ratio BR$_i$ for the $i$-th channel. The astrophysical factor $J_{\rm ann}$ and $J_{\rm dec}$ are built on the target DM density, integrated over the line of sight (l.o.s.) to the dSph and the source extension $\Delta\Omega$:
\begin{equation}\label{eqn:jfac}
J_{\rm ann}(\Delta \Omega) = \int_{\Delta \Omega}  \int_{\rm l.o.s.} \rho_{\rm DM}^2(\ell, \Omega) \;d\ell\,d\Omega %\qquad \rm{Annihilating~DM}\\
\end{equation}
\begin{equation}\label{eqn:dfac}
J_{\rm dec}(\Delta \Omega) = \int_{\Delta \Omega}  \int_{\rm l.o.s.} \rho_{\rm DM}(\ell, \Omega) \; d\ell\,d\Omega %\qquad \rm{Decaying~DM}
\end{equation}

\begin{figure}[htbp]
    \centering
    \includegraphics[width=\textwidth]{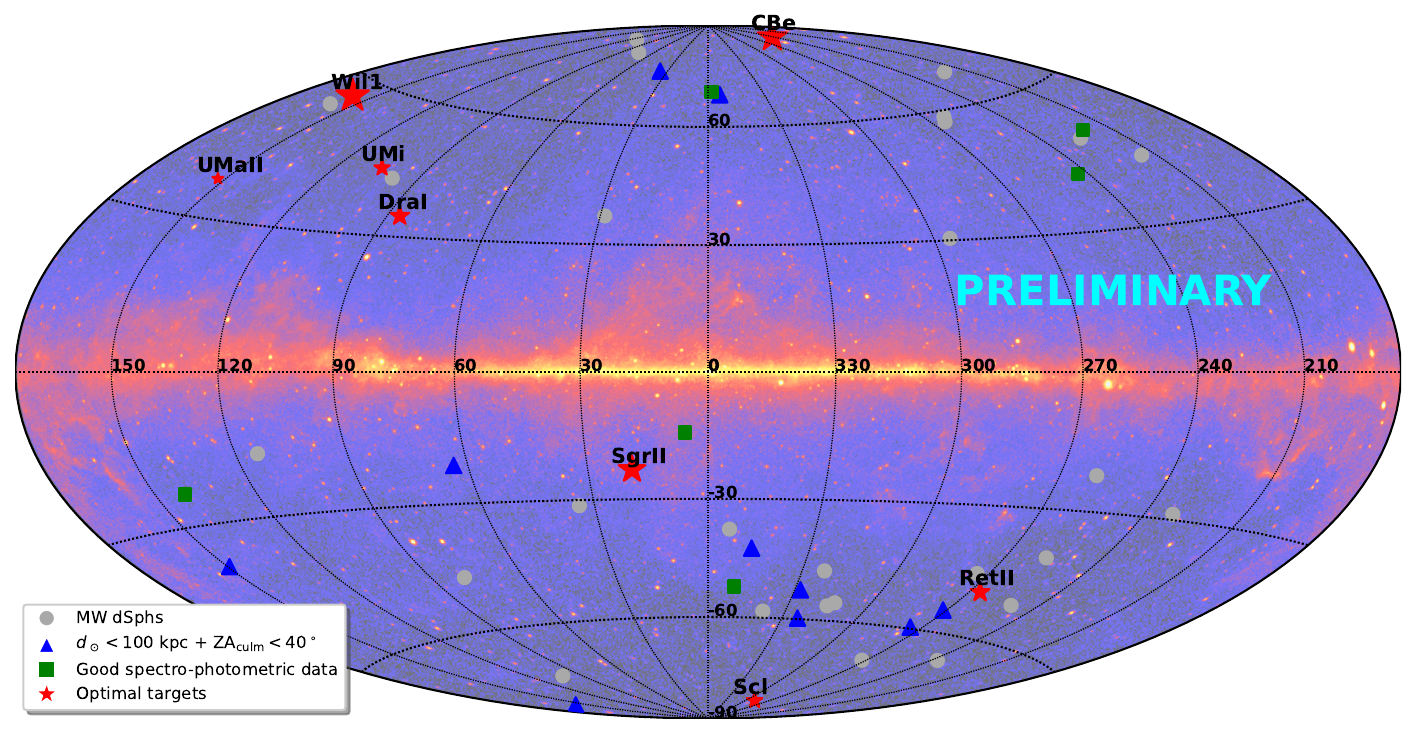}
    \caption{Sky distribution of known dSphs, superimposed to the {\itshape Fermi}-LAT $\gamma$-ray background (credits: NASA/DOE/{\itshape Fermi}-LAT Coll.). The adopted symbols correspond to targets passing incrementally stringent selection cuts (see legend). The optimal targets ({\itshape red stars}) are highlighted with symbols of increasing size, proportional to the value of their $\log{J_{\rm ann}(0^\circ.1)}$.}
    \label{fig:dsphmap}
\end{figure}

Since both $J_{\rm ann}$ and $J_{\rm dec}$ linearly increase the corresponding signal model, it is of utmost relevance to clearly assess the astrophysical factors for the DM halos of interest in order to infer DM properties (in case of a positive signal detection) or to provide robust constraints to the particle DM parameters (in case of a null detection). The literature features several methods and assumptions for deriving the DM astrophysical factors for dSphs \cite{str08,bon15,ger15,mar15,hay16}; here, we provide our own estimates based on a common framework of settings. Such novel calculations are particularly relevant to influences the choice of the optimal targets to be observed with CTA.

\begin{table}[htbp]
\begin{center} %[h!t]
\begin{tabular}{l|cc|cc}
\hline
\hline
\multicolumn{5}{l}{}\\
 & \multicolumn{2}{c}{E{\scriptsize INASTO}} & \multicolumn{2}{c}{B{\scriptsize URKERT}}\\
\multicolumn{5}{l}{}\\
\cline{2-5}
\multicolumn{5}{l}{}\\
Name & $\log{J_{\rm ann}(0^\circ.1)}$ & $\log{J_{\rm dec}(0^\circ.1)}$ & $\log{J_{\rm ann}(0^\circ.1)}$ & $\log{J_{\rm dec}(0^\circ.1)}$\\
\multicolumn{5}{l}{}\\
\hline
\multicolumn{5}{l}{}\\
Coma Berenices (CBe) & $18.7^{+0.4}_{-0.5}$ & $17.6^{+0.6}_{-0.3}$ & $18.7^{+0.6}_{-0.5}$ & $17.8^{+0.7}_{-0.5}$ \\
Draco I (DraI) & $18.3^{+0.3}_{-0.2}$ & $17.3^{+0.1}_{-0.1}$ & $18.1^{+0.1}_{-0.3}$ & $17.3^{+0.1}_{-0.1}$ \\
Reticulum II (RetII) & $18.3^{+0.3}_{-0.3}$ & $17.3^{+0.4}_{-0.3}$ & $18.4^{+0.7}_{-0.6}$ & $17.4^{+0.9}_{-0.4}$ \\
Sculptor (Scl) & $18.2^{+0.3}_{-0.2}$ & $17.2^{+0.1}_{-0.1}$ & $17.9^{+0.1}_{-0.1}$ & $17.2^{+0.1}_{-0.1}$ \\
Sagittarius II (SgrII) & $18.6^{+1.0}_{-0.8}$ & $17.4^{+0.8}_{-0.7}$ & $19.4^{+1.1}_{-1.1}$ & $18.0^{+1.0}_{-1.0}$ \\
Ursa Major II (UMaII) & $18.1^{+0.7}_{-0.7}$ & $17.3^{+0.3}_{-0.2}$ & $17.8^{+0.7}_{-0.8}$ & $17.3^{+0.7}_{-0.5}$ \\
Ursa Minor (UMi) & $18.2^{+0.1}_{-0.1}$ & $17.6^{+0.1}_{-0.1}$ & $16.2^{+0.3}_{-0.3}$ & $16.5^{+0.4}_{-0.2}$ \\
Willman 1 (Wil1) & $18.9^{+0.4}_{-0.4}$ & $17.3^{+0.4}_{-0.3}$ & $18.9^{+0.5}_{-0.4}$ & $17.4^{+0.7}_{-0.3}$ \\
\multicolumn{5}{l}{}\\
\hline
\end{tabular}
\caption{Astrophysical factors for DM annihilation and decay of the selected 8 optimal dSphs, computed at $0^\circ.1$ for both the Einasto and Burkert DM density profiles. All the values of $J_{\rm ann}$ are given in logarithmic GeV$^2$ cm$^{-5}$ and all those of $J_{\rm dec}$ in logarithmic GeV cm$^{-2}$.}
\label{tab:tab-3}
\end{center}
\end{table}

Our derivation of the dSph astrophysical factors is based on the procedure described in \cite{bon15}, that makes use of the publicly available {\footnotesize CLUMPY} code \cite{hut19}. {\footnotesize CLUMPY} allows to perform a MC dynamical analysis of the DM halos around a dSph that is treated as a steady-state, collisionless systems in spherical symmetry and with negligible rotation, in which the contribution of the stellar component to the total mass can be also neglected. Under such assumptions, the MC analysis relies on the solution of the second-order spherical Jeans equation \cite{bin08}:
\begin{equation}\label{eqn:jeans}
    \frac{1}{n^*(r)}\left\{
    \frac{d}{dr}\left[
    n^*(r)\overline{v_r^2}
    \right]
    \right\} + 2\beta_{\rm ani}(r)\frac{\overline{v_r^2}}{r} \simeq -\frac{GM_{\rm DM}(r)}{r}
\end{equation}
where $n^*(r)$ is the stellar number density, $\overline{v_r^2}$ is the square velocity dispersion and $\beta_{\rm ani}(r) = 1 - \overline{v_\theta^2}/\overline{v_r^2}$ is the velocity anisotropy of the dSph. Feeding {\footnotesize CLUMPY} with such quantities in appropriate forms -- i.e. either fixed input, sets of discrete input values over which performing the MC, or sets of free parameters that describe the adopted radial profiles \cite{bon15b} -- we are able to solve Eq. \ref{eqn:jeans} and thus obtain the DM density profiles $\rho_{\rm DM}(r)$ along the dSph radial coordinate $r$.

\begin{figure}[htbp]
    \centering
    \includegraphics[width=.8\textwidth,angle=-90]{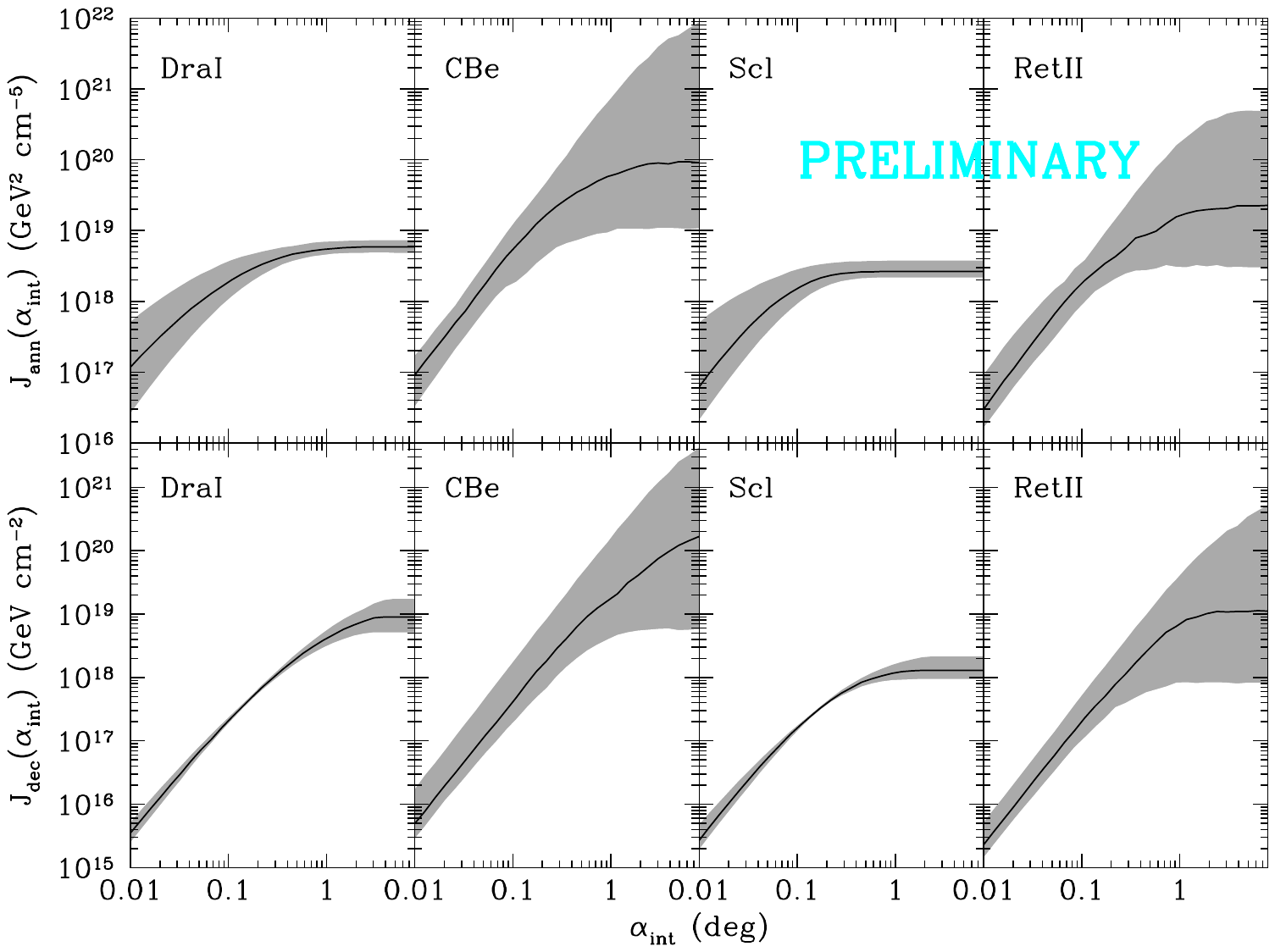}
    \caption{{\itshape Top panels:} DM annihilation astrophysical factor profiles $J_{\rm ann}(\alpha_{\rm int})$ as functions of the integration angle $\alpha_{\rm int}$ ({\itshape black solid lines}), along with the corresponding uncertainties at 68\% confidence level ({\itshape grey shaded areas}), for some of the optimal dSphs reported in Tab. \ref{tab:tab-3} (Einasto DM density profile only). {\itshape Bottom panels:} DM decay astrophysical factor profiles $J_{\rm dec}(\alpha_{\rm int})$ ({\itshape black solid lines}) and corresponding uncertainties at 68\% confidence level ({\itshape grey shaded areas}) for the same targets.}
    \label{fig:jfac}
\end{figure}

First, we select all of the dSphs whose expected DM signal is relatively strong and can be integrated over the entire CTA energy window: to this aim, we choose targets that are within 100 kpc distance and culminate at zenith angles ZA $< 40^\circ$ if observed from one of the CTA sites (La Palma for CTA North and Paranal for CTA South, respectively). Then, we restrict our calculations to those objects that have spectro-photometric data samples containing more statistics than the number of free parameters in the Jeans analysis, in order to obtain meaningful results. For such targets, we compute the expected $J_{\rm ann}$ and $J_{\rm dec}$ from the posterior distributions of the parameters describing their DM density profiles, both for a cuspy (Einasto) \cite{ein65} and a cored shape (Burkert) \cite{bur95}:
\begin{eqnarray}\label{eqn:dmprof}
\rho_{\rm DM}^{\rm (Ein)} = \rho_s \exp{\left\{
-\frac{2}{\alpha}\left[
\left(
\frac{r}{r_s}
\right)^\alpha - 1
\right]
\right\}}\\
\rho_{\rm DM}^{\rm (Bur)} = \frac{\rho_s}{\left(
1 + \frac{r}{r_s}
\right)\left[
1 + \left(
\frac{r}{r_s}
\right)^2
\right]}
\end{eqnarray}
Finally, we select as optimal dSphs for each hemisphere those sources that exhibit $J_{\rm ann}(0^\circ.1) \gtrsim 10^{18}$ GeV$^2$ cm$^{-5}$ for integration angles of $0^\circ.1$; we report the values of $J_{\rm ann}$ and $J_{\rm dec}$ for such sources in Tab. \ref{tab:tab-3}. In Fig. \ref{fig:dsphmap}, we show the sky positions in Galactic coordinates of the optimal dSphs, along with those that have not passed all of the selection cuts; in Fig. \ref{fig:jfac}, the profiles of $J_{\rm ann}$ and $J_{\rm dec}$ (Einasto DM profile only) as a function of the instrumental integration angle $\alpha_{\rm int}$ are shown for those dSphs -- Coma Berenices (CBe), Draco I (DraI), Reticulum II (RetII) and Sculptor (Scl) -- that represent the best trade-off between the expected signal intensity and the uncertainties on the astrophysical factor values.

\section{Sensitivity of CTA to $\gamma$-rays from dark matter self-interaction}
\label{sec:ctasens}

\begin{figure}[htbp]
    \centering
    \begin{minipage}{0.49\textwidth}
        \includegraphics[width=\textwidth]{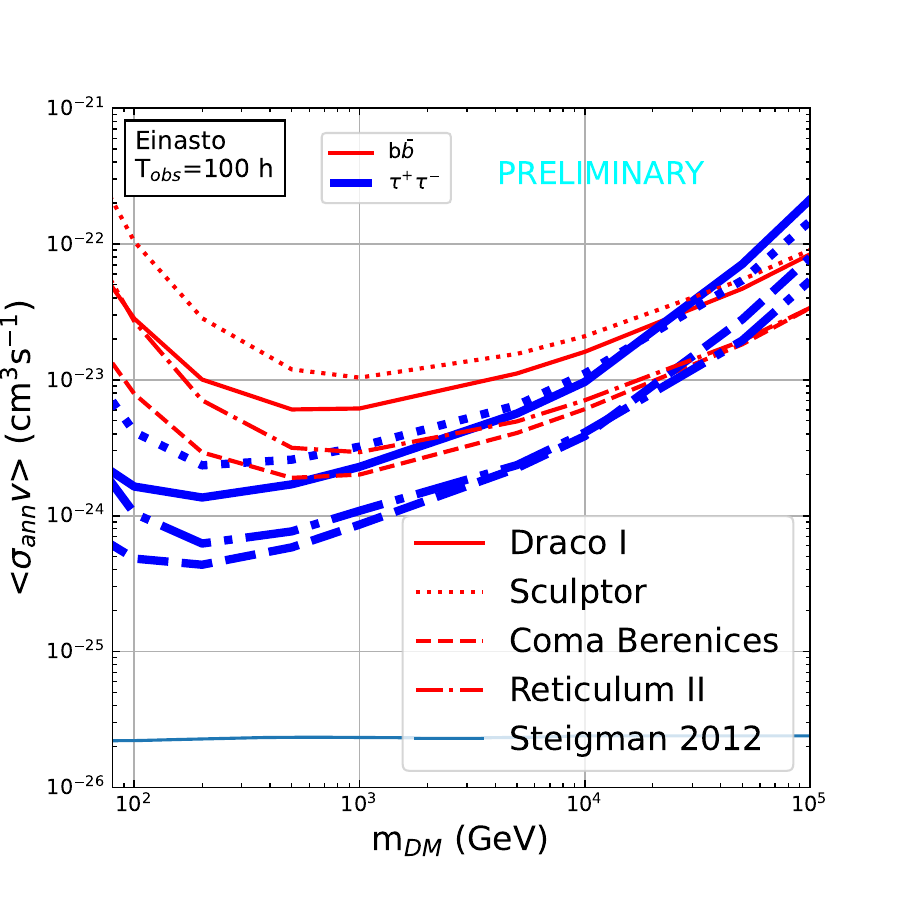}
    \end{minipage}
    \begin{minipage}{0.49\textwidth}
        \includegraphics[width=\textwidth]{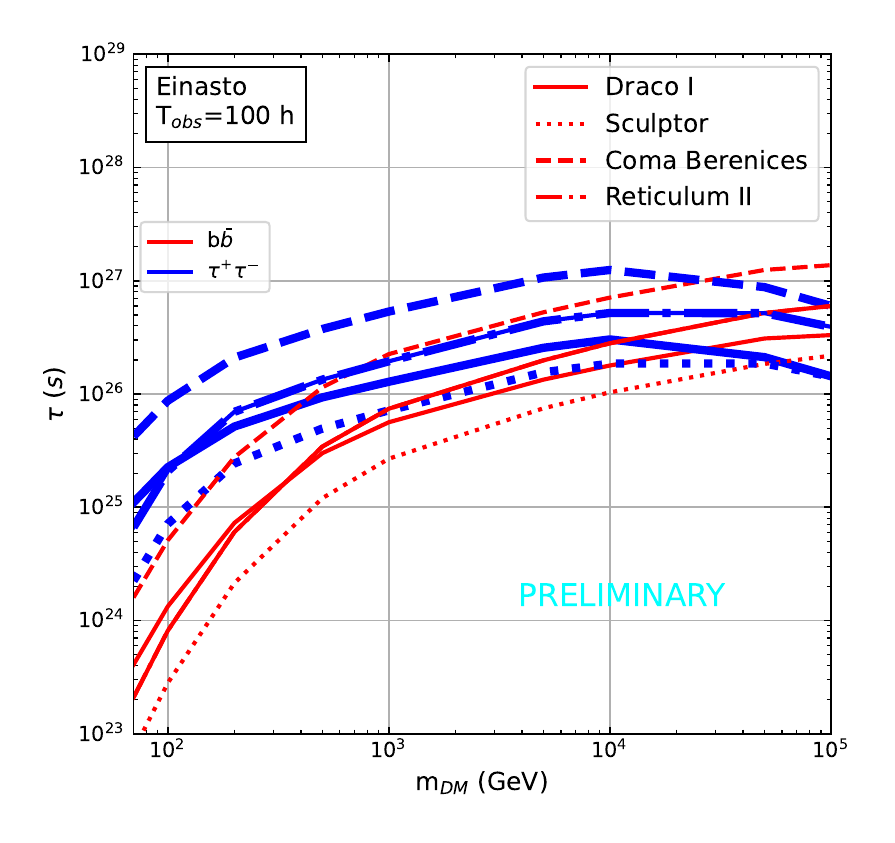}
    \end{minipage}
    \caption{{\itshape Left panel:} upper limits on annihilating DM cross sections for the four selected dSphs -- CBe ({\itshape dashed lines}), DraI ({\itshape solid lines}), RetII ({\itshape dot-dashed lines}) and Scl ({\itshape dotted lines}) -- with the Einasto DM density profile derived by {\scriptsize CLUMPY}, computed assuming 100 h of observation for annihilation in the two pure DM channels $b\bar{b}$ ({\itshape thin red lines}) and $\tau^+\tau^-$ ({\itshape thick blue lines}). The thermal-relic cross section limit \cite{ste12} ({\itshape cyan line}) is also indicated. {\itshape Right panel:} lower limits on the particle lifetime for models of DM decaying into the same channels.}
    \label{fig:dmann}
\end{figure}

We then use the astrophysical factors computed with {\footnotesize CLUMPY} for the four optimal dSphs reported above to predict the CTA sensitivity to DM signals. To this end, we make use of the official CTA analysis code {\ttfamily GammaPy} \cite{dei17} coupled with the CTA instrument response functions (IRFs)\footnote{Available at \url{https://zenodo.org/record/5499840\#.ZGTfjdZBweO}.}. For a given modeled target flux and a set of IRFs and observational parameters, {\ttfamily GammaPy} builds a maximum likelihood estimator over the parameters of interests, and estimates its uncertainty via the likelihood profiling in both spatial and energy bins. The combined likelihood for having $n_{ij}$ counts in all energy ($N_E$) and spatial bins ($N_P$) is:
\begin{equation}\label{eq:lkl_combined}
 \mathcal{L}(\vec{\alpha};\mathbf{\nu}|\mathbf{D})= \mathcal{L}\left[
 \vec{\alpha}|n_{ij}(\mathbf{\nu})
 \right] = \prod_{i=1}^{N_E}\prod_{j=1}^{N_P} \frac{\mu_{ij}^{n_{ij}}e^{-\mu_{ij}}}{n_{ij}!}
\end{equation}
where $\mu_{ij}$ is the expected Poissonian mean count for each bin, $\mathbf{\nu}$ are the nuisance parameters and $\mathbf{D}$ is the simulated data set. Since this likelihood analysis produces no significant detection of DM signals for any target, we compute the CTA sensitivity to annihilation cross sections $\langle \sigma v \rangle$ and decay lifetimes $\tau$ over the range of DM particle masses $0.1 - 100$ TeV.

\begin{figure}[htbp]
    \centering
    \includegraphics[width=\textwidth]{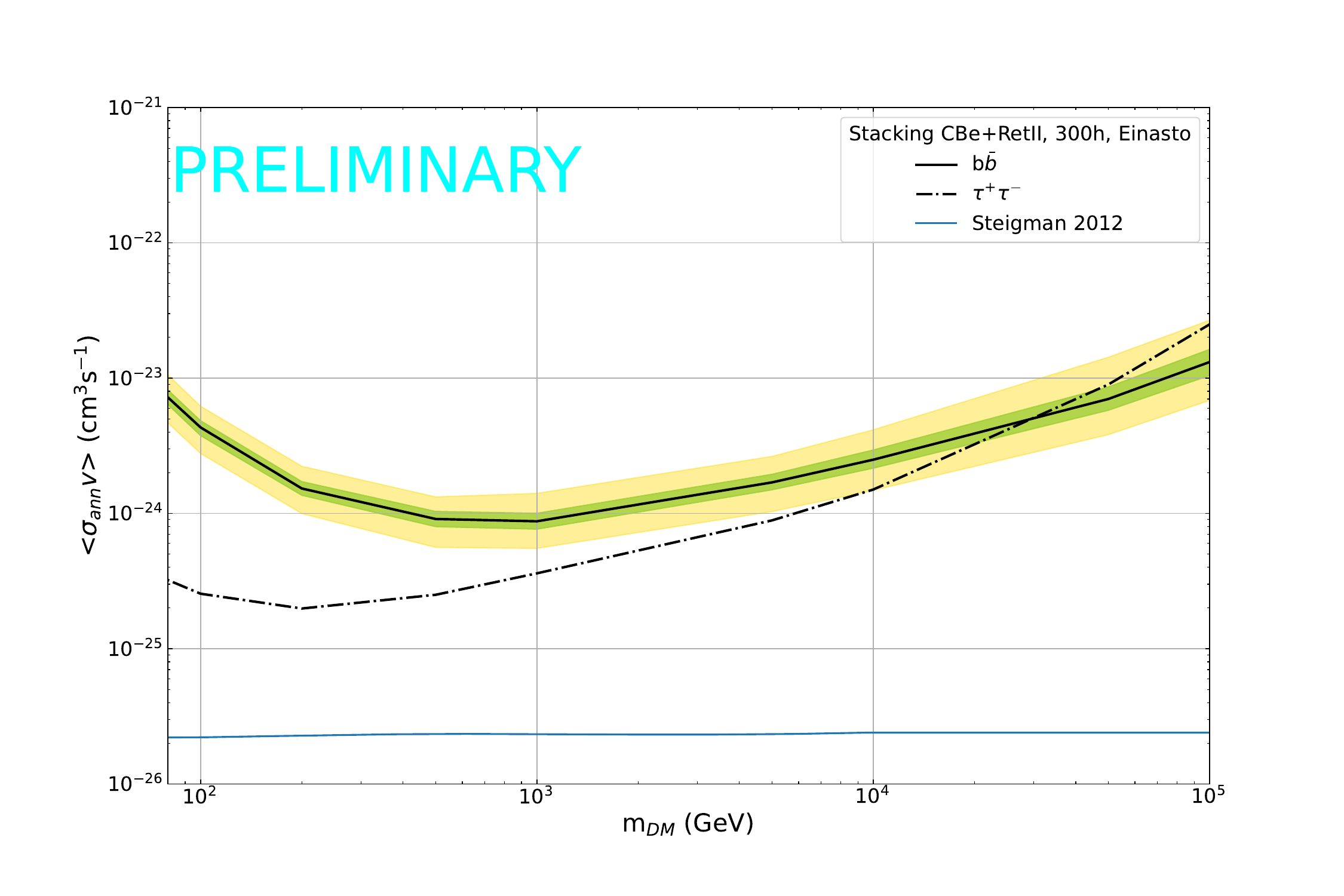}
    \caption{Combined limits from the two best dSphs -- CBe ({\itshape solid line}) and RetII ({\itshape dot-dashed line}) -- observed for 300 h each for both the $b\bar{b}$ and $\tau^+\tau^-$ channels, with the uncertainty on the stacked cross-section limit due to photon statistics at 68\% ({\itshape green band}) and 95\% confidence level ({\itshape yellow band}) reported for the $b\bar{b}$ channel.}
    \label{fig:dmstack}
\end{figure}

In Fig. \ref{fig:dmann} we report the upper limits to $\langle \sigma v \rangle$ and lower limits to $\tau$ as a function of $m_{\rm DM}$ for the dSphs with the highest astrophysical factors according to Tab. \ref{tab:tab-3} (Einasto DM density profile only), assuming 100\% annihilation in either the $b\bar{b}$ or $\tau^+\tau^-$ SM channels. Since CTA will invest a total of $\sim$500-h observing time on the DM searches in dSphs, likely observing each target for $\gtrsim$100 h, we also compute the prospects for the scenario in which we combine $\sim$600-h observations of the two overall best targets, showing them in Fig. \ref{fig:dmstack} for the DM annihilation case. All of the obtained cross-section limits sit between $\mathcal{O}(10^{-24}) - \mathcal{O}(10^{-22})$ cm$^3$ s$^{-1}$ for the $b\bar{b}$ channel and $\mathcal{O}(10^{-25}) - \mathcal{O}(10^{-23})$ cm$^3$ s$^{-1}$ for the $\tau^+\tau^-$ channel, whereas the lifetime limits may exceed $\gtrsim$10$^{27}$ s for $m_{\rm DM} \gtrsim 10$ TeV.

\section{Summary and outlook}
\label{sec:sum}

%The results obtained above allows us to discuss the observational strategy of CTA toward the class of dSphs. 

This work presents the limits on the particle DM parameter space that CTA can obtain in the search for signals from WIMP annihilation or decay in the halos around dSphs. Since these constraints are strongly affected by uncertainties related to the determination of the DM amount in dSphs based on the astronomical knowledge of their stellar content, the programming of deep spectro-photometric surveys on presently known objects or targets of future discovery is of paramount importance to reduce such biases and thus obtain a final set of optimal dSphs to be pointed at.

Also, not all the dSphs residing in the MW halo have already been discovered. Based on $N$-body simulations carried out in the framework of the Aquarius Project \cite{spr08}, we expect $124^{+40}_{-27}$ satellite galaxies with $V$-band absolute magnitude $M_V > 0$ in a MW-like halo \cite{new17}; only roughly half of this number of dSphs has presently been discovered. Since the detection of fainter targets requires surveys that go $\sim$4 mag deeper than the current ones, this task will be accessible to future facilities like the ``Vera C. Rubin Observatory'' \cite{ive19} to potentially provide up to $\sim$60 new dSph candidates. Such considerations highlight further the relevance of future campaigns aimed at characterizing the stellar population of dSphs, in order to expand the sample of potential CTA targets and achieve more accurate determinations of their astrophysical factors. 

\acknowledgments % equivalent to \section*{ACKNOWLEDGMENTS}       
{\footnotesize This work was conducted in the context of the CTA DMEP Working Group. We gratefully acknowledge financial support from the agencies and organizations listed here: \url{https://www.cta-observatory.org/consortium\_acknowledgments}. FGS acknowledges financial support from the PRIN MIUR project ``ASTRI/CTA Data Challenge'' (PI: P. Caraveo), contract 298/2017. {\footnotesize CLUMPY} is licensed under the GNU General Public License (GPLv2). This research has made use of {\ttfamily GammaPy} (\url{https://www.gammapy.org}), a community-developed core Python package for TeV $\gamma$-ray astronomy, and of the CTA instrument response functions (version {\ttfamily prod5-v0.1}) provided by the CTA Consortium and Observatory (see \url{https://www.cta-observatory.org/science/ctao-performance/} for more details).}

%% Full authors list (ONLY FOR COLLABORATIONS)
\clearpage
\section*{Full Authors List: CTA Consortium}
%
%\noindent \textbf{Note comment afterwards:} Collaborations have the possibility to provide an authors list in xml format which will be used while generating the DOI entries making the full authors list searchable in databases like Inspire HEP. For instructions please go to icrc2021.desy.de/proceedings or contact us under icrc2021proc@desy.de.\\
%
\scriptsize
%\noindent
%\documentclass[a4paper,10pt]{article}

%\begin{document}
\pagestyle{empty}
\centering\LARGE
The CTA Consortium\\[0.5cm]
\normalsize
\raggedright
  \mbox{K.~Abe$^{\ref{AFFIL::JapanUTokai}}$}, 
  \mbox{S.~Abe$^{\ref{AFFIL::JapanUTokyoICRR}}$\orcid{0000-0001-7250-3596}}, 
  \mbox{A.~Acharyya$^{\ref{AFFIL::USAUAlabamaTuscaloosa}}$\orcid{0000-0002-2028-9230}}, 
  \mbox{R.~Adam$^{\ref{AFFIL::FranceOCotedAzur},\ref{AFFIL::FranceLLREcolePolytechnique}}$}, 
  \mbox{A.~Aguasca-Cabot$^{\ref{AFFIL::SpainICCUB}}$\orcid{0000-0001-8816-4920}}, 
  \mbox{I.~Agudo$^{\ref{AFFIL::SpainIAACSIC}}$\orcid{0000-0002-3777-6182}}, 
  \mbox{J.~Alfaro$^{\ref{AFFIL::ChileUPontificiaCatolicadeChile}}$}, 
  \mbox{N.~Alvarez-Crespo$^{\ref{AFFIL::SpainUCMAltasEnergias}}$}, 
  \mbox{R.~Alves~Batista$^{\ref{AFFIL::SpainIFTUAMCSIC}}$\orcid{0000-0003-2656-064X}}, 
  \mbox{J.-P.~Amans$^{\ref{AFFIL::FranceObservatoiredeParis}}$}, 
  \mbox{E.~Amato$^{\ref{AFFIL::ItalyOArcetri}}$\orcid{0000-0002-9881-8112}}, 
  \mbox{F.~Ambrosino$^{\ref{AFFIL::ItalyORoma}}$}, 
  \mbox{E.~O.~Ang\"uner$^{\ref{AFFIL::TurkeyTubitak}}$\orcid{0000-0002-4712-4292}}, 
  \mbox{L.~A.~Antonelli$^{\ref{AFFIL::ItalyORoma}}$\orcid{0000-0002-5037-9034}}, 
  \mbox{C.~Aramo$^{\ref{AFFIL::ItalyINFNNapoli}}$}, 
  \mbox{C.~Arcaro$^{\ref{AFFIL::ItalyINFNPadova}}$\orcid{0000-0002-1998-9707}}, 
  \mbox{L.~Arrabito$^{\ref{AFFIL::FranceLUPMUMontpellier}}$\orcid{0000-0003-4727-7288}}, 
  \mbox{K.~Asano$^{\ref{AFFIL::JapanUTokyoICRR}}$}, 
  \mbox{J.~Aschersleben$^{\ref{AFFIL::NetherlandsUGroningen}}$\orcid{0000-0002-6097-7898}}, 
  \mbox{H.~Ashkar$^{\ref{AFFIL::FranceLLREcolePolytechnique}}$\orcid{0000-0002-2153-1818}}, 
  \mbox{L.~Augusto~Stuani$^{\ref{AFFIL::BrazilIFSCUSaoPaulo}}$}, 
  \mbox{D.~Baack$^{\ref{AFFIL::GermanyUDortmundTU}}$\orcid{0000-0002-2311-4460}}, 
  \mbox{M.~Backes$^{\ref{AFFIL::NamibiaUNamibia},\ref{AFFIL::SouthAfricaNWU}}$\orcid{0000-0002-9326-6400}}, 
  \mbox{C.~Balazs$^{\ref{AFFIL::AustraliaUMonash}}$\orcid{0000-0001-7154-1726}}, 
  \mbox{M.~Balbo$^{\ref{AFFIL::SwitzerlandUGenevaISDC}}$\orcid{0000-0002-6556-3344}}, 
  \mbox{A.~Baquero~Larriva$^{\ref{AFFIL::SpainUCMAltasEnergias},\ref{AFFIL::EcuadorUAzuay}}$\orcid{0000-0002-1757-5826}}, 
  \mbox{V.~Barbosa~Martins$^{\ref{AFFIL::GermanyDESY}}$\orcid{0000-0002-5085-8828}}, 
  \mbox{U.~Barres~de~Almeida$^{\ref{AFFIL::BrazilCBPF},\ref{AFFIL::BrazilIAGUSaoPaulo}}$\orcid{0000-0001-7909-588X}}, 
  \mbox{J.~A.~Barrio$^{\ref{AFFIL::SpainUCMAltasEnergias}}$\orcid{0000-0002-0965-0259}}, 
  \mbox{D.~Bastieri$^{\ref{AFFIL::ItalyUPadovaandINFN}}$\orcid{0000-0002-6954-8862}}, 
  \mbox{P.~I.~Batista$^{\ref{AFFIL::GermanyDESY}}$\orcid{0000-0001-8138-1391}}, 
  \mbox{I.~Batkovic$^{\ref{AFFIL::ItalyUPadovaandINFN}}$\orcid{0000-0002-1209-2542}}, 
  \mbox{R.~Batzofin$^{\ref{AFFIL::GermanyUPotsdam}}$\orcid{0000-0002-5797-3386}}, 
  \mbox{J.~Baxter$^{\ref{AFFIL::JapanUTokyoICRR}}$\orcid{0009-0004-9545-794X}}, 
  \mbox{G.~Beck$^{\ref{AFFIL::SouthAfricaUWitwatersrand}}$\orcid{0000-0003-4916-4914}}, 
  \mbox{J.~Becker~Tjus$^{\ref{AFFIL::GermanyUBochum}}$\orcid{0000-0002-1748-7367}}, 
  \mbox{L.~Beiske$^{\ref{AFFIL::GermanyUDortmundTU}}$\orcid{0009-0004-7097-0122}}, 
  \mbox{D.~Belardinelli$^{\ref{AFFIL::ItalyINFNRomaTorVergata}}$\orcid{0000-0001-9332-5733}}, 
  \mbox{W.~Benbow$^{\ref{AFFIL::USACfAHarvardSmithsonian}}$\orcid{0000-0003-2098-170X}}, 
  \mbox{E.~Bernardini$^{\ref{AFFIL::ItalyUPadovaandINFN}}$\orcid{0000-0003-3108-1141}}, 
  \mbox{J.~Bernete~Medrano$^{\ref{AFFIL::SpainCIEMAT}}$\orcid{0000-0002-8108-7552}}, 
  \mbox{K.~Bernl\"ohr$^{\ref{AFFIL::GermanyMPIK}}$\orcid{0000-0001-8065-3252}}, 
  \mbox{A.~Berti$^{\ref{AFFIL::GermanyMPP}}$\orcid{0000-0003-0396-4190}}, 
  \mbox{V.~Beshley$^{\ref{AFFIL::UkraineIAPMMLviv}}$}, 
  \mbox{P.~Bhattacharjee$^{\ref{AFFIL::FranceLAPPUSavoieMontBlanc}}$\orcid{0000-0002-0258-3831}}, 
  \mbox{S.~Bhattacharyya$^{\ref{AFFIL::SloveniaUNovaGoricaCAC}}$\orcid{0000-0002-6569-5953}}, 
  \mbox{B.~Bi$^{\ref{AFFIL::GermanyIAAT}}$\orcid{0000-0003-0455-4038}}, 
  \mbox{N.~Biederbeck$^{\ref{AFFIL::GermanyUDortmundTU}}$\orcid{0000-0003-3708-9785}}, 
  \mbox{A.~Biland$^{\ref{AFFIL::SwitzerlandETHZurich}}$}, 
  \mbox{E.~Bissaldi$^{\ref{AFFIL::ItalyPolitecnicoBari},\ref{AFFIL::ItalyINFNBari}}$\orcid{0000-0001-9935-8106}}, 
  \mbox{O.~Blanch$^{\ref{AFFIL::SpainIFAEBIST}}$\orcid{0000-0002-8380-1633}}, 
  \mbox{J.~Blazek$^{\ref{AFFIL::CzechRepublicFZU}}$\orcid{0000-0002-5870-8947}}, 
  \mbox{C.~Boisson$^{\ref{AFFIL::FranceObservatoiredeParis}}$\orcid{0000-0001-5893-1797}}, 
  \mbox{J.~Bolmont$^{\ref{AFFIL::FranceLPNHEUSorbonne}}$\orcid{0000-0003-4739-8389}}, 
  \mbox{G.~Bonnoli$^{\ref{AFFIL::ItalyOBrera},\ref{AFFIL::ItalyINFNPisa}}$\orcid{0000-0003-2464-9077}}, 
  \mbox{P.~Bordas$^{\ref{AFFIL::SpainICCUB}}$}, 
  \mbox{Z.~Bosnjak$^{\ref{AFFIL::CroatiaUZagreb}}$\orcid{0000-0001-6536-0320}}, 
  \mbox{F.~Bradascio$^{\ref{AFFIL::FranceCEAIRFUDPhP}}$\orcid{0000-0002-7750-5256}}, 
  \mbox{C.~Braiding$^{\ref{AFFIL::AustraliaUAdelaide}}$}, 
  \mbox{E.~Bronzini$^{\ref{AFFIL::ItalyOASBologna}}$\orcid{0000-0001-8378-4303}}, 
  \mbox{R.~Brose$^{\ref{AFFIL::IrelandDIAS}}$}, 
  \mbox{A.~M.~Brown$^{\ref{AFFIL::UnitedKingdomUDurham}}$\orcid{0000-0003-0259-3148}}, 
  \mbox{F.~Brun$^{\ref{AFFIL::FranceCEAIRFUDPhP}}$}, 
  \mbox{G.~Brunelli$^{\ref{AFFIL::ItalyOASBologna},\ref{AFFIL::SpainIAACSIC}}$}, 
  \mbox{A.~Bulgarelli$^{\ref{AFFIL::ItalyOASBologna}}$}, 
  \mbox{I.~Burelli$^{\ref{AFFIL::ItalyUUdineandINFNTrieste}}$}, 
  \mbox{L.~Burmistrov$^{\ref{AFFIL::SwitzerlandUGenevaDPNC}}$}, 
  \mbox{M.~Burton$^{\ref{AFFIL::UnitedKingdomArmaghObservatoryandPlanetarium},\ref{AFFIL::AustraliaUNewSouthWales}}$\orcid{0000-0001-7289-1998}}, 
  \mbox{T.~Bylund$^{\ref{AFFIL::FranceCEAIRFUDAp}}$\orcid{0000-0003-2946-1313}}, 
  \mbox{P.~G.~Calisse$^{\ref{AFFIL::GermanyCTAOHeidelberg}}$}, 
  \mbox{A.~Campoy-Ordaz$^{\ref{AFFIL::SpainUABandCERESIEEC}}$\orcid{0000-0001-9352-8936}}, 
  \mbox{B.~K.~Cantlay$^{\ref{AFFIL::ThailandUKasetsart},\ref{AFFIL::ThailandNARIT}}$\orcid{0009-0002-8750-6401}}, 
  \mbox{M.~Capalbi$^{\ref{AFFIL::ItalyIASFPalermo}}$\orcid{0000-0002-9558-2394}}, 
  \mbox{A.~Caproni$^{\ref{AFFIL::BrazilUCruzeirodoSul}}$\orcid{0000-0001-9707-3895}}, 
  \mbox{R.~Capuzzo-Dolcetta$^{\ref{AFFIL::ItalyORoma}}$\orcid{0000-0002-6871-9519}}, 
  \mbox{C.~Carlile$^{\ref{AFFIL::SwedenLundObservatory}}$}, 
  \mbox{S.~Caroff$^{\ref{AFFIL::FranceLAPPUSavoieMontBlanc}}$\orcid{0000-0002-1103-130X}}, 
  \mbox{A.~Carosi$^{\ref{AFFIL::ItalyORoma}}$}, 
  \mbox{R.~Carosi$^{\ref{AFFIL::ItalyINFNPisa}}$}, 
  \mbox{M.-S.~Carrasco$^{\ref{AFFIL::FranceCPPMUAixMarseille}}$}, 
  \mbox{E.~Cascone$^{\ref{AFFIL::ItalyOCapodimonte}}$\orcid{0000-0002-7425-7517}}, 
  \mbox{F.~Cassol$^{\ref{AFFIL::FranceCPPMUAixMarseille}}$\orcid{0000-0002-0372-1992}}, 
  \mbox{N.~Castrejon$^{\ref{AFFIL::SpainUAlcala}}$}, 
  \mbox{F.~Catalani$^{\ref{AFFIL::BrazilEELUSaoPaulo}}$}, 
  \mbox{D.~Cerasole$^{\ref{AFFIL::ItalyUandINFNBari}}$\orcid{0000-0003-2033-756X}}, 
  \mbox{M.~Cerruti$^{\ref{AFFIL::FranceAPCUParisCite}}$\orcid{0000-0001-7891-699X}}, 
  \mbox{S.~Chaty$^{\ref{AFFIL::FranceAPCUParisCite}}$\orcid{0000-0002-5769-8601}}, 
  \mbox{A.~W.~Chen$^{\ref{AFFIL::SouthAfricaUWitwatersrand}}$\orcid{0000-0001-6425-5692}}, 
  \mbox{M.~Chernyakova$^{\ref{AFFIL::IrelandDCU}}$\orcid{0000-0002-9735-3608}}, 
  \mbox{A.~Chiavassa$^{\ref{AFFIL::ItalyINFNTorino},\ref{AFFIL::ItalyUTorino}}$\orcid{0000-0001-6183-2589}}, 
  \mbox{J.~Chudoba$^{\ref{AFFIL::CzechRepublicFZU}}$\orcid{0000-0002-6425-2579}}, 
  \mbox{C.~H.~Coimbra~Araujo$^{\ref{AFFIL::BrazilUFPR}}$}, 
  \mbox{V.~Conforti$^{\ref{AFFIL::ItalyOASBologna}}$\orcid{0000-0002-0007-3520}}, 
  \mbox{F.~Conte$^{\ref{AFFIL::GermanyMPIK}}$\orcid{0000-0002-3083-8539}}, 
  \mbox{J.~L.~Contreras$^{\ref{AFFIL::SpainUCMAltasEnergias}}$\orcid{0000-0001-7282-2394}}, 
  \mbox{C.~Cossou$^{\ref{AFFIL::FranceCEAIRFUDAp}}$\orcid{0000-0001-5350-4796}}, 
  \mbox{A.~Costa$^{\ref{AFFIL::ItalyOCatania}}$\orcid{0000-0003-0344-8911}}, 
  \mbox{H.~Costantini$^{\ref{AFFIL::FranceCPPMUAixMarseille}}$}, 
  \mbox{P.~Cristofari$^{\ref{AFFIL::FranceObservatoiredeParis}}$}, 
  \mbox{O.~Cuevas$^{\ref{AFFIL::ChileUdeValparaiso}}$}, 
  \mbox{Z.~Curtis-Ginsberg$^{\ref{AFFIL::USAUWisconsin}}$\orcid{0000-0002-0194-7576}}, 
  \mbox{G.~D'Amico$^{\ref{AFFIL::NorwayUBergen}}$}, 
  \mbox{F.~D'Ammando$^{\ref{AFFIL::ItalyRadioastronomiaINAF}}$\orcid{0000-0001-7618-7527}}, 
  \mbox{M.~Dadina$^{\ref{AFFIL::ItalyOASBologna}}$\orcid{0000-0002-7858-7564}}, 
  \mbox{M.~Dalchenko$^{\ref{AFFIL::SwitzerlandUGenevaDPNC}}$\orcid{0000-0002-0137-136X}}, 
  \mbox{L.~David$^{\ref{AFFIL::GermanyDESY}}$\orcid{0000-0003-2341-9261}}, 
  \mbox{I.~D.~Davids$^{\ref{AFFIL::NamibiaUNamibia}}$\orcid{0000-0002-6476-964X}}, 
  \mbox{F.~Dazzi$^{\ref{AFFIL::ItalyINAF}}$\orcid{0000-0001-5409-6544}}, 
  \mbox{A.~De~Angelis$^{\ref{AFFIL::ItalyUPadovaandINFN}}$}, 
  \mbox{M.~de~Bony~de~Lavergne$^{\ref{AFFIL::FranceCEAIRFUDAp}}$\orcid{0000-0002-4650-1666}}, 
  \mbox{V.~De~Caprio$^{\ref{AFFIL::ItalyOCapodimonte}}$\orcid{0000-0002-4587-8963}}, 
  \mbox{G.~De~Cesare$^{\ref{AFFIL::ItalyOASBologna}}$\orcid{0000-0003-0869-7183}}, 
  \mbox{E.~M.~de~Gouveia~Dal~Pino$^{\ref{AFFIL::BrazilIAGUSaoPaulo}}$\orcid{0000-0001-8058-4752}}, 
  \mbox{B.~De~Lotto$^{\ref{AFFIL::ItalyUUdineandINFNTrieste}}$\orcid{0000-0003-3624-4480}}, 
  \mbox{M.~De~Lucia$^{\ref{AFFIL::ItalyINFNNapoli}}$\orcid{0000-0002-0519-9149}}, 
  \mbox{R.~de~Menezes$^{\ref{AFFIL::ItalyINFNTorino},\ref{AFFIL::ItalyUTorino}}$\orcid{0000-0001-5489-4925}}, 
  \mbox{M.~de~Naurois$^{\ref{AFFIL::FranceLLREcolePolytechnique}}$\orcid{0000-0002-7245-201X}}, 
  \mbox{E.~de~Ona~Wilhelmi$^{\ref{AFFIL::GermanyDESY}}$\orcid{0000-0002-5401-0744}}, 
  \mbox{N.~De~Simone$^{\ref{AFFIL::GermanyDESY}}$}, 
  \mbox{V.~de~Souza$^{\ref{AFFIL::BrazilIFSCUSaoPaulo}}$}, 
  \mbox{L.~del~Peral$^{\ref{AFFIL::SpainUAlcala}}$}, 
  \mbox{M.~V.~del~Valle$^{\ref{AFFIL::BrazilIAGUSaoPaulo}}$\orcid{0000-0002-5444-0795}}, 
  \mbox{E.~Delagnes$^{\ref{AFFIL::FranceCEAIRFUDEDIP}}$}, 
  \mbox{A.~G.~Delgado~Giler$^{\ref{AFFIL::BrazilIFSCUSaoPaulo},\ref{AFFIL::NetherlandsUGroningen}}$\orcid{0000-0003-2190-9857}}, 
  \mbox{C.~Delgado$^{\ref{AFFIL::SpainCIEMAT}}$\orcid{0000-0002-7014-4101}}, 
  \mbox{M.~Dell'aiera$^{\ref{AFFIL::FranceLAPPUSavoieMontBlanc}}$\orcid{0000-0002-5221-0240}}, 
  \mbox{R.~Della~Ceca$^{\ref{AFFIL::ItalyOBrera}}$\orcid{0000-0001-7551-2252}}, 
  \mbox{M.~Della~Valle$^{\ref{AFFIL::ItalyOCapodimonte}}$}, 
  \mbox{D.~della~Volpe$^{\ref{AFFIL::SwitzerlandUGenevaDPNC}}$\orcid{0000-0001-8530-7447}}, 
  \mbox{D.~Depaoli$^{\ref{AFFIL::GermanyMPIK}}$\orcid{0000-0002-2672-4141}}, 
  \mbox{A.~Dettlaff$^{\ref{AFFIL::GermanyMPP}}$}, 
  \mbox{T.~Di~Girolamo$^{\ref{AFFIL::ItalyUNapoli},\ref{AFFIL::ItalyINFNNapoli}}$\orcid{0000-0003-2339-4471}}, 
  \mbox{A.~Di~Piano$^{\ref{AFFIL::ItalyOASBologna}}$}, 
  \mbox{F.~Di~Pierro$^{\ref{AFFIL::ItalyINFNTorino}}$\orcid{0000-0003-4861-432X}}, 
  \mbox{R.~Di~Tria$^{\ref{AFFIL::ItalyUandINFNBari}}$\orcid{0009-0007-1088-5307}}, 
  \mbox{L.~Di~Venere$^{\ref{AFFIL::ItalyINFNBari}}$}, 
  \mbox{C.~D{\'\i}az-Bahamondes$^{\ref{AFFIL::ChileUPontificiaCatolicadeChile}}$}, 
  \mbox{C.~Dib$^{\ref{AFFIL::ChileUTecnicaFedericoSantaMaria}}$\orcid{0000-0003-4146-906X}}, 
  \mbox{S.~Diebold$^{\ref{AFFIL::GermanyIAAT}}$\orcid{0000-0002-8042-2443}}, 
  \mbox{R.~Dima$^{\ref{AFFIL::ItalyUPadovaandINFN}}$}, 
  \mbox{A.~Dinesh$^{\ref{AFFIL::SpainUCMAltasEnergias}}$}, 
  \mbox{A.~Djannati-Ata{\"\i}$^{\ref{AFFIL::FranceAPCUParisCite}}$\orcid{0000-0002-4924-1708}}, 
  \mbox{J.~Djuvsland$^{\ref{AFFIL::NorwayUBergen}}$\orcid{0000-0002-6488-8219}}, 
  \mbox{A.~Dom{\'\i}nguez$^{\ref{AFFIL::SpainUCMAltasEnergias}}$}, 
  \mbox{R.~M.~Dominik$^{\ref{AFFIL::GermanyUDortmundTU}}$\orcid{0000-0003-4168-7200}}, 
  \mbox{A.~Donini$^{\ref{AFFIL::ItalyORoma}}$\orcid{0000-0002-3066-724X}}, 
  \mbox{D.~Dorner$^{\ref{AFFIL::GermanyUWurzburg},\ref{AFFIL::SwitzerlandETHZurich}}$\orcid{0000-0001-8823-479X}}, 
  \mbox{J.~D\"orner$^{\ref{AFFIL::GermanyUBochum}}$\orcid{0000-0001-6692-6293}}, 
  \mbox{M.~Doro$^{\ref{AFFIL::ItalyUPadovaandINFN}}$\orcid{0000-0001-9104-3214}}, 
  \mbox{R.~D.~C.~dos~Anjos$^{\ref{AFFIL::BrazilUFPR}}$\orcid{0000-0002-6463-2272}}, 
  \mbox{J.-L.~Dournaux$^{\ref{AFFIL::FranceObservatoiredeParis}}$}, 
  \mbox{D.~Dravins$^{\ref{AFFIL::SwedenLundObservatory}}$\orcid{0000-0001-9024-0400}}, 
  \mbox{C.~Duangchan$^{\ref{AFFIL::GermanyUErlangenECAP},\ref{AFFIL::ThailandNARIT}}$\orcid{0009-0003-8227-6552}}, 
  \mbox{C.~Dubos$^{\ref{AFFIL::FranceIJCLab}}$}, 
  \mbox{L.~Ducci$^{\ref{AFFIL::GermanyIAAT}}$}, 
  \mbox{V.~V.~Dwarkadas$^{\ref{AFFIL::USAUChicagoDAA}}$\orcid{0000-0002-4661-7001}}, 
  \mbox{J.~Ebr$^{\ref{AFFIL::CzechRepublicFZU}}$}, 
  \mbox{C.~Eckner$^{\ref{AFFIL::FranceLAPPUSavoieMontBlanc},\ref{AFFIL::FranceLAPTh}}$}, 
  \mbox{K.~Egberts$^{\ref{AFFIL::GermanyUPotsdam}}$}, 
  \mbox{S.~Einecke$^{\ref{AFFIL::AustraliaUAdelaide}}$\orcid{0000-0001-9687-8237}}, 
  \mbox{D.~Els\"asser$^{\ref{AFFIL::GermanyUDortmundTU}}$\orcid{0000-0001-6796-3205}}, 
  \mbox{G.~Emery$^{\ref{AFFIL::FranceCPPMUAixMarseille}}$\orcid{0000-0001-6155-4742}}, 
  \mbox{M.~Escobar~Godoy$^{\ref{AFFIL::USASCIPP}}$}, 
  \mbox{J.~Escudero$^{\ref{AFFIL::SpainIAACSIC}}$\orcid{0000-0002-4131-655X}}, 
  \mbox{P.~Esposito$^{\ref{AFFIL::ItalyIUSSPaviaINAF},\ref{AFFIL::ItalyIASFMilano}}$\orcid{0000-0003-4849-5092}}, 
  \mbox{D.~Falceta-Goncalves$^{\ref{AFFIL::BrazilEACHUSaoPaulo}}$\orcid{0000-0002-1914-6654}}, 
  \mbox{V.~Fallah~Ramazani$^{\ref{AFFIL::GermanyUBochum}}$\orcid{0000-0001-8991-7744}}, 
  \mbox{A.~Faure$^{\ref{AFFIL::FranceLUPMUMontpellier}}$}, 
  \mbox{E.~Fedorova$^{\ref{AFFIL::ItalyORoma},\ref{AFFIL::UkraineAstObsofUKyiv}}$\orcid{0000-0002-8882-7496}}, 
  \mbox{S.~Fegan$^{\ref{AFFIL::FranceLLREcolePolytechnique}}$\orcid{0000-0002-9978-2510}}, 
  \mbox{K.~Feijen$^{\ref{AFFIL::FranceAPCUParisCite}}$\orcid{0000-0003-1476-3714}}, 
  \mbox{Q.~Feng$^{\ref{AFFIL::USACfAHarvardSmithsonian}}$}, 
  \mbox{G.~Ferrand$^{\ref{AFFIL::CanadaUManitoba},\ref{AFFIL::JapanRIKEN}}$\orcid{0000-0002-4231-8717}}, 
  \mbox{F.~Ferrarotto$^{\ref{AFFIL::ItalyINFNRomaLaSapienza}}$\orcid{0000-0001-5464-0378}}, 
  \mbox{E.~Fiandrini$^{\ref{AFFIL::ItalyUPerugiaandINFN}}$}, 
  \mbox{A.~Fiasson$^{\ref{AFFIL::FranceLAPPUSavoieMontBlanc}}$}, 
  \mbox{V.~Fioretti$^{\ref{AFFIL::ItalyOASBologna}}$\orcid{0000-0002-6082-5384}}, 
  \mbox{L.~Foffano$^{\ref{AFFIL::ItalyIAPS}}$\orcid{0000-0002-0709-9707}}, 
  \mbox{L.~Font~Guiteras$^{\ref{AFFIL::SpainUABandCERESIEEC}}$\orcid{0000-0003-2109-5961}}, 
  \mbox{G.~Fontaine$^{\ref{AFFIL::FranceLLREcolePolytechnique}}$\orcid{0000-0002-6443-5025}}, 
  \mbox{S.~Fr\"ose$^{\ref{AFFIL::GermanyUDortmundTU}}$\orcid{0000-0003-1832-4129}}, 
  \mbox{S.~Fukami$^{\ref{AFFIL::SwitzerlandETHZurich}}$}, 
  \mbox{Y.~Fukui$^{\ref{AFFIL::JapanUNagoya}}$\orcid{0000-0002-8966-9856}}, 
  \mbox{S.~Funk$^{\ref{AFFIL::GermanyUErlangenECAP}}$\orcid{0000-0002-2012-0080}}, 
  \mbox{D.~Gaggero$^{\ref{AFFIL::ItalyINFNPisa}}$}, 
  \mbox{G.~Galanti$^{\ref{AFFIL::ItalyIASFMilano}}$\orcid{0000-0001-7254-3029}}, 
  \mbox{G.~Galaz$^{\ref{AFFIL::ChileUPontificiaCatolicadeChile}}$\orcid{0000-0002-8835-0739}}, 
  \mbox{Y.~A.~Gallant$^{\ref{AFFIL::FranceLUPMUMontpellier}}$}, 
  \mbox{S.~Gallozzi$^{\ref{AFFIL::ItalyORoma}}$\orcid{0000-0003-4456-9875}}, 
  \mbox{V.~Gammaldi$^{\ref{AFFIL::SpainIFTUAMCSIC}}$\orcid{0000-0003-1826-6117}}, 
  \mbox{C.~Gasbarra$^{\ref{AFFIL::ItalyINFNRomaTorVergata}}$\orcid{0000-0001-8335-9614}}, 
  \mbox{M.~Gaug$^{\ref{AFFIL::SpainUABandCERESIEEC}}$\orcid{0000-0001-8442-7877}}, 
  \mbox{A.~Ghalumyan$^{\ref{AFFIL::ArmeniaNSLAlikhanyan}}$}, 
  \mbox{F.~Gianotti$^{\ref{AFFIL::ItalyOASBologna}}$\orcid{0000-0003-4666-119X}}, 
  \mbox{M.~Giarrusso$^{\ref{AFFIL::ItalyINFNCatania}}$}, 
  \mbox{N.~Giglietto$^{\ref{AFFIL::ItalyPolitecnicoBari},\ref{AFFIL::ItalyINFNBari}}$\orcid{0000-0002-9021-2888}}, 
  \mbox{F.~Giordano$^{\ref{AFFIL::ItalyUandINFNBari}}$\orcid{0000-0002-8651-2394}}, 
  \mbox{A.~Giuliani$^{\ref{AFFIL::ItalyIASFMilano}}$}, 
  \mbox{J.-F.~Glicenstein$^{\ref{AFFIL::FranceCEAIRFUDPhP}}$}, 
  \mbox{J.~Glombitza$^{\ref{AFFIL::GermanyUErlangenECAP}}$}, 
  \mbox{P.~Goldoni$^{\ref{AFFIL::FranceAPCUParisCiteCEAaffiliatedpersonnel}}$\orcid{0000-0001-5638-5817}}, 
  \mbox{J.~M.~Gonz\'alez$^{\ref{AFFIL::ChileUAndresBello}}$\orcid{0000-0002-2413-0681}}, 
  \mbox{M.~M.~Gonz\'alez$^{\ref{AFFIL::MexicoUNAMMexico}}$}, 
  \mbox{J.~Goulart~Coelho$^{\ref{AFFIL::BrazilUFES}}$\orcid{0000-0001-9386-1042}}, 
  \mbox{J.~Granot$^{\ref{AFFIL::IsraelOpenUniversityofIsrael},\ref{AFFIL::USAGWUWashingtonDC}}$}, 
  \mbox{D.~Grasso$^{\ref{AFFIL::ItalyINFNPisa}}$}, 
  \mbox{R.~Grau$^{\ref{AFFIL::SpainIFAEBIST}}$\orcid{0000-0002-1891-6290}}, 
  \mbox{D.~Green$^{\ref{AFFIL::GermanyMPP}}$\orcid{0000-0003-0768-2203}}, 
  \mbox{J.~G.~Green$^{\ref{AFFIL::GermanyMPP}}$\orcid{0000-0002-1130-6692}}, 
  \mbox{T.~Greenshaw$^{\ref{AFFIL::UnitedKingdomULiverpool}}$}, 
  \mbox{G.~Grolleron$^{\ref{AFFIL::FranceLPNHEUSorbonne}}$}, 
  \mbox{J.~Grube$^{\ref{AFFIL::UnitedKingdomKingsCollege}}$}, 
  \mbox{O.~Gueta$^{\ref{AFFIL::GermanyDESY}}$\orcid{0000-0002-9440-2398}}, 
  \mbox{S.~Gunji$^{\ref{AFFIL::JapanUYamagata}}$\orcid{0000-0002-5881-2445}}, 
  \mbox{D.~Hadasch$^{\ref{AFFIL::JapanUTokyoICRR}}$\orcid{0000-0001-8663-6461}}, 
  \mbox{P.~Hamal$^{\ref{AFFIL::CzechRepublicFZU}}$\orcid{0000-0003-3139-7234}}, 
  \mbox{W.~Hanlon$^{\ref{AFFIL::USACfAHarvardSmithsonian}}$\orcid{0000-0002-0109-4737}}, 
  \mbox{S.~Hara$^{\ref{AFFIL::JapanUYamanashiGakuin}}$\orcid{0009-0001-1220-7717}}, 
  \mbox{V.~M.~Harvey$^{\ref{AFFIL::AustraliaUAdelaide}}$\orcid{0000-0001-9090-8415}}, 
  \mbox{K.~Hashiyama$^{\ref{AFFIL::JapanUTokyoICRR}}$}, 
  \mbox{T.~Hassan$^{\ref{AFFIL::SpainCIEMAT}}$\orcid{0000-0002-4758-9196}}, 
  \mbox{M.~Heller$^{\ref{AFFIL::SwitzerlandUGenevaDPNC}}$}, 
  \mbox{S.~Hern\'andez~Cadena$^{\ref{AFFIL::MexicoUNAMMexico}}$\orcid{0000-0002-2565-8365}}, 
  \mbox{J.~Hie$^{\ref{AFFIL::FranceIRAPUToulouse}}$}, 
  \mbox{N.~Hiroshima$^{\ref{AFFIL::JapanUTokyoICRR}}$}, 
  \mbox{B.~Hnatyk$^{\ref{AFFIL::UkraineAstObsofUKyiv}}$\orcid{0000-0001-7113-4709}}, 
  \mbox{R.~Hnatyk$^{\ref{AFFIL::UkraineAstObsofUKyiv}}$}, 
  \mbox{D.~Hoffmann$^{\ref{AFFIL::FranceCPPMUAixMarseille}}$\orcid{0000-0001-5209-5265}}, 
  \mbox{W.~Hofmann$^{\ref{AFFIL::GermanyMPIK}}$}, 
  \mbox{M.~Holler$^{\ref{AFFIL::AustriaUInnsbruck}}$}, 
  \mbox{D.~Horan$^{\ref{AFFIL::FranceLLREcolePolytechnique}}$}, 
  \mbox{P.~Horvath$^{\ref{AFFIL::CzechRepublicUOlomouc}}$\orcid{0000-0002-6710-5339}}, 
  \mbox{T.~Hovatta$^{\ref{AFFIL::FinlandUTurku}}$}, 
  \mbox{D.~Hrupec$^{\ref{AFFIL::CroatiaUOsijek}}$\orcid{0000-0002-7027-5021}}, 
  \mbox{S.~Hussain$^{\ref{AFFIL::BrazilIAGUSaoPaulo},\ref{AFFIL::ItalyGSSIandINFNAquila}}$\orcid{0000-0002-0458-0490}}, 
  \mbox{M.~Iarlori$^{\ref{AFFIL::ItalyUandINFNAquila}}$}, 
  \mbox{T.~Inada$^{\ref{AFFIL::JapanUTokyoICRR}}$\orcid{0000-0002-6923-9314}}, 
  \mbox{F.~Incardona$^{\ref{AFFIL::ItalyOCatania}}$}, 
  \mbox{Y.~Inome$^{\ref{AFFIL::JapanUTokyoICRR}}$}, 
  \mbox{S.~Inoue$^{\ref{AFFIL::JapanRIKEN}}$}, 
  \mbox{F.~Iocco$^{\ref{AFFIL::ItalyUNapoli},\ref{AFFIL::ItalyINFNNapoli}}$}, 
  \mbox{K.~Ishio$^{\ref{AFFIL::PolandULodz}}$}, 
  \mbox{M.~Jamrozy$^{\ref{AFFIL::PolandUJagiellonian}}$\orcid{0000-0002-0870-7778}}, 
  \mbox{P.~Janecek$^{\ref{AFFIL::CzechRepublicFZU}}$}, 
  \mbox{F.~Jankowsky$^{\ref{AFFIL::GermanyLSW}}$}, 
  \mbox{C.~Jarnot$^{\ref{AFFIL::FranceIRAPUToulouse}}$}, 
  \mbox{P.~Jean$^{\ref{AFFIL::FranceIRAPUToulouse}}$\orcid{0000-0002-1757-9560}}, 
  \mbox{I.~Jim\'enez~Mart{\'\i}nez$^{\ref{AFFIL::SpainCIEMAT}}$\orcid{0000-0003-2150-6919}}, 
  \mbox{W.~Jin$^{\ref{AFFIL::USAUAlabamaTuscaloosa}}$\orcid{0000-0002-1089-1754}}, 
  \mbox{L.~Jocou$^{\ref{AFFIL::FranceIPAGUGrenobleAlpes}}$}, 
  \mbox{C.~Juramy-Gilles$^{\ref{AFFIL::FranceLPNHEUSorbonne}}$\orcid{0000-0002-3145-9258}}, 
  \mbox{J.~Jurysek$^{\ref{AFFIL::CzechRepublicFZU}}$\orcid{0000-0002-3130-4168}}, 
  \mbox{O.~Kalekin$^{\ref{AFFIL::GermanyUErlangenECAP}}$}, 
  \mbox{D.~Kantzas$^{\ref{AFFIL::FranceLAPTh}}$\orcid{0000-0002-7364-606X}}, 
  \mbox{V.~Karas$^{\ref{AFFIL::CzechRepublicASU}}$}, 
  \mbox{S.~Kaufmann$^{\ref{AFFIL::UnitedKingdomUDurham}}$}, 
  \mbox{D.~Kerszberg$^{\ref{AFFIL::SpainIFAEBIST}}$\orcid{0000-0002-5289-1509}}, 
  \mbox{B.~Kh\'elifi$^{\ref{AFFIL::FranceAPCUParisCite}}$\orcid{0000-0001-6876-5577}}, 
  \mbox{D.~B.~Kieda$^{\ref{AFFIL::USAUUtah}}$\orcid{0000-0003-4785-0101}}, 
  \mbox{T.~Kleiner$^{\ref{AFFIL::GermanyDESY}}$\orcid{0000-0002-4260-9186}}, 
  \mbox{W.~Klu\'zniak$^{\ref{AFFIL::PolandNicolausCopernicusAstronomicalCenter}}$}, 
  \mbox{Y.~Kobayashi$^{\ref{AFFIL::JapanUTokyoICRR}}$}, 
  \mbox{K.~Kohri$^{\ref{AFFIL::JapanKEK}}$}, 
  \mbox{N.~Komin$^{\ref{AFFIL::SouthAfricaUWitwatersrand}}$\orcid{0000-0003-3280-0582}}, 
  \mbox{P.~Kornecki$^{\ref{AFFIL::FranceObservatoiredeParis}}$\orcid{0000-0002-2706-7438}}, 
  \mbox{K.~Kosack$^{\ref{AFFIL::FranceCEAIRFUDAp}}$\orcid{0000-0001-8424-3621}}, 
  \mbox{H.~Kubo$^{\ref{AFFIL::JapanUTokyoICRR}}$\orcid{0000-0001-9159-9853}}, 
  \mbox{J.~Kushida$^{\ref{AFFIL::JapanUTokai}}$\orcid{0000-0002-8002-8585}}, 
  \mbox{A.~La~Barbera$^{\ref{AFFIL::ItalyIASFPalermo}}$\orcid{0000-0002-5880-8913}}, 
  \mbox{N.~La~Palombara$^{\ref{AFFIL::ItalyIASFMilano}}$\orcid{0000-0001-7015-6359}}, 
  \mbox{M.~L\'ainez$^{\ref{AFFIL::SpainUCMAltasEnergias}}$\orcid{0000-0003-3848-922X}}, 
  \mbox{A.~Lamastra$^{\ref{AFFIL::ItalyORoma}}$\orcid{0000-0003-2403-913X}}, 
  \mbox{J.~Lapington$^{\ref{AFFIL::UnitedKingdomULeicester}}$}, 
  \mbox{S.~Lazarevi\'c$^{\ref{AFFIL::AustraliaUWesternSydney}}$\orcid{0000-0001-6109-8548}}, 
  \mbox{J.~Lazendic-Galloway$^{\ref{AFFIL::AustraliaUMonash}}$}, 
  \mbox{S.~Leach$^{\ref{AFFIL::UnitedKingdomULeicester}}$\orcid{0000-0003-2129-3175}}, 
  \mbox{M.~Lemoine-Goumard$^{\ref{AFFIL::FranceLP2IUBordeaux}}$}, 
  \mbox{J.-P.~Lenain$^{\ref{AFFIL::FranceLPNHEUSorbonne}}$\orcid{0000-0001-7284-9220}}, 
  \mbox{G.~Leto$^{\ref{AFFIL::ItalyOCatania}}$\orcid{0000-0002-0040-5011}}, 
  \mbox{F.~Leuschner$^{\ref{AFFIL::GermanyIAAT}}$\orcid{0000-0001-9037-0272}}, 
  \mbox{E.~Lindfors$^{\ref{AFFIL::FinlandUTurku}}$}, 
  \mbox{M.~Linhoff$^{\ref{AFFIL::GermanyUDortmundTU}}$\orcid{0000-0001-7993-8189}}, 
  \mbox{I.~Liodakis$^{\ref{AFFIL::FinlandUTurku}}$\orcid{0000-0001-9200-4006}}, 
  \mbox{L.~Lo{\"\i}c$^{\ref{AFFIL::FranceCEAIRFUDPhP}}$}, 
  \mbox{S.~Lombardi$^{\ref{AFFIL::ItalyORoma}}$\orcid{0000-0002-6336-865X}}, 
  \mbox{F.~Longo$^{\ref{AFFIL::ItalyUandINFNTrieste}}$\orcid{0000-0003-2501-2270}}, 
  \mbox{R.~L\'opez-Coto$^{\ref{AFFIL::SpainIAACSIC}}$}, 
  \mbox{M.~L\'opez-Moya$^{\ref{AFFIL::SpainUCMAltasEnergias}}$\orcid{0000-0002-8791-7908}}, 
  \mbox{A.~L\'opez-Oramas$^{\ref{AFFIL::SpainIAC}}$\orcid{0000-0003-4603-1884}}, 
  \mbox{S.~Loporchio$^{\ref{AFFIL::ItalyPolitecnicoBari},\ref{AFFIL::ItalyINFNBari}}$}, 
  \mbox{J.~Lozano~Bahilo$^{\ref{AFFIL::SpainUAlcala}}$\orcid{0000-0003-0613-140X}}, 
  \mbox{P.~L.~Luque-Escamilla$^{\ref{AFFIL::SpainUJaen}}$}, 
  \mbox{O.~Macias$^{\ref{AFFIL::NetherlandsUAmsterdam}}$\orcid{0000-0001-8867-2693}}, 
  \mbox{G.~Maier$^{\ref{AFFIL::GermanyDESY}}$\orcid{0000-0001-9868-4700}}, 
  \mbox{P.~Majumdar$^{\ref{AFFIL::IndiaSahaInstitute}}$\orcid{0000-0002-5481-5040}}, 
  \mbox{D.~Malyshev$^{\ref{AFFIL::GermanyIAAT}}$\orcid{0000-0001-9689-2194}}, 
  \mbox{D.~Malyshev$^{\ref{AFFIL::GermanyUErlangenECAP}}$\orcid{0000-0002-9102-4854}}, 
  \mbox{D.~Mandat$^{\ref{AFFIL::CzechRepublicFZU}}$}, 
  \mbox{G.~Manic\`o$^{\ref{AFFIL::ItalyINFNCatania},\ref{AFFIL::ItalyUCatania}}$}, 
  \mbox{P.~Marinos$^{\ref{AFFIL::AustraliaUAdelaide}}$\orcid{0000-0003-1734-0215}}, 
  \mbox{S.~Markoff$^{\ref{AFFIL::NetherlandsUAmsterdam}}$\orcid{0000-0001-9564-0876}}, 
  \mbox{I.~M\'arquez$^{\ref{AFFIL::SpainIAACSIC}}$\orcid{0000-0003-2629-1945}}, 
  \mbox{P.~Marquez$^{\ref{AFFIL::SpainIFAEBIST}}$\orcid{0000-0002-9591-7967}}, 
  \mbox{G.~Marsella$^{\ref{AFFIL::ItalyUPalermo},\ref{AFFIL::ItalyINFNCatania}}$\orcid{0000-0002-3152-8874}}, 
  \mbox{J.~Mart{\'\i}$^{\ref{AFFIL::SpainUJaen}}$}, 
  \mbox{P.~Martin$^{\ref{AFFIL::FranceIRAPUToulouse}}$\orcid{0000-0002-7670-6320}}, 
  \mbox{G.~A.~Mart{\'\i}nez$^{\ref{AFFIL::SpainCIEMAT}}$\orcid{0000-0002-1061-8520}}, 
  \mbox{M.~Mart{\'\i}nez$^{\ref{AFFIL::SpainIFAEBIST}}$}, 
  \mbox{O.~Martinez$^{\ref{AFFIL::SpainUCMElectronica},\ref{AFFIL::SpainUPCMadrid}}$\orcid{0000-0002-3353-7707}}, 
  \mbox{C.~Marty$^{\ref{AFFIL::FranceIRAPUToulouse}}$}, 
  \mbox{A.~Mas-Aguilar$^{\ref{AFFIL::SpainUCMAltasEnergias}}$\orcid{0000-0002-8893-9009}}, 
  \mbox{M.~Mastropietro$^{\ref{AFFIL::ItalyORoma}}$\orcid{0000-0002-6324-5713}}, 
  \mbox{G.~Maurin$^{\ref{AFFIL::FranceLAPPUSavoieMontBlanc}}$}, 
  \mbox{W.~Max-Moerbeck$^{\ref{AFFIL::ChileUdeChile}}$\orcid{0000-0002-5491-5244}}, 
  \mbox{D.~Mazin$^{\ref{AFFIL::JapanUTokyoICRR},\ref{AFFIL::GermanyMPP}}$}, 
  \mbox{D.~Melkumyan$^{\ref{AFFIL::GermanyDESY}}$}, 
  \mbox{S.~Menchiari$^{\ref{AFFIL::ItalyOArcetri},\ref{AFFIL::ItalyINFNPisa}}$}, 
  \mbox{E.~Mestre$^{\ref{AFFIL::SpainICECSIC}}$}, 
  \mbox{J.-L.~Meunier$^{\ref{AFFIL::FranceLPNHEUSorbonne}}$}, 
  \mbox{D.~M.-A.~Meyer$^{\ref{AFFIL::GermanyUPotsdam}}$\orcid{0000-0001-8258-9813}}, 
  \mbox{D.~Miceli$^{\ref{AFFIL::ItalyINFNPadova}}$\orcid{0000-0002-2686-0098}}, 
  \mbox{M.~Michailidis$^{\ref{AFFIL::GermanyIAAT}}$}, 
  \mbox{J.~Micha{\l}owski$^{\ref{AFFIL::PolandIFJ}}$}, 
  \mbox{T.~Miener$^{\ref{AFFIL::SpainUCMAltasEnergias}}$}, 
  \mbox{J.~M.~Miranda$^{\ref{AFFIL::SpainUCMElectronica},\ref{AFFIL::SpainIPARCOSInstitute}}$}, 
  \mbox{A.~Mitchell$^{\ref{AFFIL::GermanyUErlangenECAP}}$\orcid{0000-0003-3631-5648}}, 
  \mbox{M.~Mizote$^{\ref{AFFIL::JapanUKonan}}$}, 
  \mbox{T.~Mizuno$^{\ref{AFFIL::JapanHASC}}$}, 
  \mbox{R.~Moderski$^{\ref{AFFIL::PolandNicolausCopernicusAstronomicalCenter}}$\orcid{0000-0002-8663-3882}}, 
  \mbox{L.~Mohrmann$^{\ref{AFFIL::GermanyMPIK}}$\orcid{0000-0002-9667-8654}}, 
  \mbox{M.~Molero$^{\ref{AFFIL::SpainIAC}}$\orcid{0000-0003-0967-715X}}, 
  \mbox{C.~Molfese$^{\ref{AFFIL::ItalyINAF}}$\orcid{0000-0002-2756-9075}}, 
  \mbox{E.~Molina$^{\ref{AFFIL::SpainIAC}}$\orcid{0000-0003-1204-5516}}, 
  \mbox{T.~Montaruli$^{\ref{AFFIL::SwitzerlandUGenevaDPNC}}$}, 
  \mbox{A.~Moralejo$^{\ref{AFFIL::SpainIFAEBIST}}$}, 
  \mbox{D.~Morcuende$^{\ref{AFFIL::SpainUCMAltasEnergias},\ref{AFFIL::SpainIAACSIC}}$\orcid{0000-0001-9400-0922}}, 
  \mbox{K.~Morik$^{\ref{AFFIL::GermanyUDortmundTU}}$\orcid{0000-0003-1153-5986}}, 
  \mbox{A.~Morselli$^{\ref{AFFIL::ItalyINFNRomaTorVergata}}$\orcid{0000-0002-7704-9553}}, 
  \mbox{E.~Moulin$^{\ref{AFFIL::FranceCEAIRFUDPhP}}$\orcid{0000-0003-4007-0145}}, 
  \mbox{V.~Moya~Zamanillo$^{\ref{AFFIL::SpainUCMAltasEnergias}}$\orcid{0000-0001-9407-5545}}, 
  \mbox{R.~Mukherjee$^{\ref{AFFIL::USABarnardCollegeColumbiaUniversity}}$\orcid{0000-0002-3223-0754}}, 
  \mbox{K.~Munari$^{\ref{AFFIL::ItalyOCatania}}$}, 
  \mbox{A.~Muraczewski$^{\ref{AFFIL::PolandNicolausCopernicusAstronomicalCenter}}$}, 
  \mbox{H.~Muraishi$^{\ref{AFFIL::JapanUKitasato}}$\orcid{0000-0003-3054-5725}}, 
  \mbox{T.~Nakamori$^{\ref{AFFIL::JapanUYamagata}}$\orcid{0000-0002-7308-2356}}, 
  \mbox{L.~Nava$^{\ref{AFFIL::ItalyOBrera}}$\orcid{0000-0001-5960-0455}}, 
  \mbox{A.~Nayak$^{\ref{AFFIL::UnitedKingdomUDurham}}$}, 
  \mbox{R.~Nemmen$^{\ref{AFFIL::BrazilIAGUSaoPaulo},\ref{AFFIL::USAStanford}}$\orcid{0000-0003-3956-0331}}, 
  \mbox{L.~Nickel$^{\ref{AFFIL::GermanyUDortmundTU}}$\orcid{0000-0001-7110-0533}}, 
  \mbox{J.~Niemiec$^{\ref{AFFIL::PolandIFJ}}$\orcid{0000-0001-6036-8569}}, 
  \mbox{D.~Nieto$^{\ref{AFFIL::SpainUCMAltasEnergias}}$\orcid{0000-0003-3343-0755}}, 
  \mbox{M.~Nievas~Rosillo$^{\ref{AFFIL::SpainIAC}}$\orcid{0000-0002-8321-9168}}, 
  \mbox{M.~Niko{\l}ajuk$^{\ref{AFFIL::PolandUBiaystok}}$\orcid{0000-0003-4075-6745}}, 
  \mbox{K.~Nishijima$^{\ref{AFFIL::JapanUTokai}}$\orcid{0000-0002-1830-4251}}, 
  \mbox{K.~Noda$^{\ref{AFFIL::JapanUTokyoICRR}}$\orcid{0000-0003-1397-6478}}, 
  \mbox{D.~Nosek$^{\ref{AFFIL::CzechRepublicUPrague}}$\orcid{0000-0001-6219-200X}}, 
  \mbox{B.~Novosyadlyj$^{\ref{AFFIL::UkraineAstObsofULviv}}$}, 
  \mbox{V.~Novotny$^{\ref{AFFIL::CzechRepublicUPrague}}$\orcid{0000-0002-4319-4541}}, 
  \mbox{S.~Nozaki$^{\ref{AFFIL::GermanyMPP}}$\orcid{0000-0002-6246-2767}}, 
  \mbox{P.~O'Brien$^{\ref{AFFIL::UnitedKingdomULeicester}}$\orcid{0000-0002-5128-1899}}, 
  \mbox{M.~Ohishi$^{\ref{AFFIL::JapanUTokyoICRR}}$\orcid{0000-0002-5056-0968}}, 
  \mbox{Y.~Ohtani$^{\ref{AFFIL::JapanUTokyoICRR}}$\orcid{0000-0001-7042-4958}}, 
  \mbox{A.~Okumura$^{\ref{AFFIL::JapanUNagoyaISEE},\ref{AFFIL::JapanUNagoyaKMI}}$\orcid{0000-0002-3055-7964}}, 
  \mbox{J.-F.~Olive$^{\ref{AFFIL::FranceIRAPUToulouse}}$}, 
  \mbox{B.~Olmi$^{\ref{AFFIL::ItalyOPalermo},\ref{AFFIL::ItalyOArcetri}}$}, 
  \mbox{R.~A.~Ong$^{\ref{AFFIL::USAUCLA}}$\orcid{0000-0002-4837-5253}}, 
  \mbox{M.~Orienti$^{\ref{AFFIL::ItalyRadioastronomiaINAF}}$\orcid{0000-0003-4470-7094}}, 
  \mbox{R.~Orito$^{\ref{AFFIL::JapanUTokushima}}$}, 
  \mbox{M.~Orlandini$^{\ref{AFFIL::ItalyOASBologna}}$\orcid{0000-0003-0946-3151}}, 
  \mbox{E.~Orlando$^{\ref{AFFIL::ItalyUandINFNTrieste}}$}, 
  \mbox{M.~Ostrowski$^{\ref{AFFIL::PolandUJagiellonian}}$\orcid{0000-0002-9199-7031}}, 
  \mbox{N.~Otte$^{\ref{AFFIL::USAGeorgiaTech}}$\orcid{0000-0002-5955-6383}}, 
  \mbox{I.~Oya$^{\ref{AFFIL::GermanyCTAOHeidelberg}}$\orcid{0000-0002-3881-9324}}, 
  \mbox{I.~Pagano$^{\ref{AFFIL::ItalyOCatania}}$\orcid{0000-0001-9573-4928}}, 
  \mbox{A.~Pagliaro$^{\ref{AFFIL::ItalyIASFPalermo}}$\orcid{0000-0002-6841-1362}}, 
  \mbox{M.~Palatiello$^{\ref{AFFIL::ItalyUUdineandINFNTrieste}}$}, 
  \mbox{G.~Panebianco$^{\ref{AFFIL::ItalyOASBologna}}$\orcid{0000-0002-3410-8613}}, 
  \mbox{J.~M.~Paredes$^{\ref{AFFIL::SpainICCUB}}$}, 
  \mbox{N.~Parmiggiani$^{\ref{AFFIL::ItalyOASBologna}}$\orcid{0000-0002-4535-5329}}, 
  \mbox{S.~R.~Patel$^{\ref{AFFIL::FranceIJCLab}}$\orcid{0000-0001-8965-7292}}, 
  \mbox{B.~Patricelli$^{\ref{AFFIL::ItalyORoma},\ref{AFFIL::ItalyUPisa}}$\orcid{0000-0001-6709-0969}}, 
  \mbox{D.~Pavlovi\'c$^{\ref{AFFIL::CroatiaURijeka}}$}, 
  \mbox{A.~Pe'er$^{\ref{AFFIL::GermanyMPP}}$\orcid{0000-0001-8667-0889}}, 
  \mbox{M.~Pech$^{\ref{AFFIL::CzechRepublicFZU}}$}, 
  \mbox{M.~Pecimotika$^{\ref{AFFIL::CroatiaURijeka},\ref{AFFIL::CroatiaIRB}}$\orcid{0000-0002-4699-1845}}, 
  \mbox{M.~Peresano$^{\ref{AFFIL::ItalyUTorino},\ref{AFFIL::ItalyINFNTorino}}$\orcid{0000-0002-7537-7334}}, 
  \mbox{J.~P\'erez-Romero$^{\ref{AFFIL::SpainIFTUAMCSIC},\ref{AFFIL::SloveniaUNovaGoricaCAC}}$\orcid{0000-0002-9408-3120}}, 
  \mbox{G.~Peron$^{\ref{AFFIL::FranceAPCUParisCite}}$}, 
  \mbox{M.~Persic$^{\ref{AFFIL::ItalyOPadova},\ref{AFFIL::ItalyOandINFNTrieste}}$\orcid{0000-0003-1853-4900}}, 
  \mbox{P.-O.~Petrucci$^{\ref{AFFIL::FranceIPAGUGrenobleAlpes}}$\orcid{0000-0001-6061-3480}}, 
  \mbox{O.~Petruk$^{\ref{AFFIL::UkraineIAPMMLviv}}$\orcid{0000-0003-3487-0349}}, 
  \mbox{F.~Pfeifle$^{\ref{AFFIL::GermanyUWurzburg}}$}, 
  \mbox{F.~Pintore$^{\ref{AFFIL::ItalyIASFPalermo}}$\orcid{0000-0002-3869-2925}}, 
  \mbox{G.~Pirola$^{\ref{AFFIL::GermanyMPP}}$}, 
  \mbox{C.~Pittori$^{\ref{AFFIL::ItalyORoma}}$\orcid{0000-0001-6661-9779}}, 
  \mbox{C.~Plard$^{\ref{AFFIL::FranceLAPPUSavoieMontBlanc}}$\orcid{0000-0002-4061-3800}}, 
  \mbox{F.~Podobnik$^{\ref{AFFIL::ItalyUSienaandINFN}}$\orcid{0000-0001-6125-9487}}, 
  \mbox{M.~Pohl$^{\ref{AFFIL::GermanyUPotsdam},\ref{AFFIL::GermanyDESY}}$\orcid{0000-0001-7861-1707}}, 
  \mbox{E.~Pons$^{\ref{AFFIL::FranceLAPPUSavoieMontBlanc}}$\orcid{0000-0002-7601-9811}}, 
  \mbox{E.~Prandini$^{\ref{AFFIL::ItalyUPadovaandINFN}}$\orcid{0000-0003-4502-9053}}, 
  \mbox{J.~Prast$^{\ref{AFFIL::FranceLAPPUSavoieMontBlanc}}$}, 
  \mbox{G.~Principe$^{\ref{AFFIL::ItalyUandINFNTrieste}}$}, 
  \mbox{C.~Priyadarshi$^{\ref{AFFIL::SpainIFAEBIST}}$\orcid{0000-0002-9160-9617}}, 
  \mbox{N.~Produit$^{\ref{AFFIL::SwitzerlandUGenevaISDC}}$\orcid{0000-0001-7138-7677}}, 
  \mbox{D.~Prokhorov$^{\ref{AFFIL::NetherlandsUAmsterdam}}$}, 
  \mbox{E.~Pueschel$^{\ref{AFFIL::GermanyDESY}}$\orcid{0000-0002-0529-1973}}, 
  \mbox{G.~P\"uhlhofer$^{\ref{AFFIL::GermanyIAAT}}$}, 
  \mbox{M.~L.~Pumo$^{\ref{AFFIL::ItalyUCatania},\ref{AFFIL::ItalyINFNCatania}}$}, 
  \mbox{M.~Punch$^{\ref{AFFIL::FranceAPCUParisCite}}$\orcid{0000-0002-4710-2165}}, 
  \mbox{A.~Quirrenbach$^{\ref{AFFIL::GermanyLSW}}$}, 
  \mbox{S.~Rain\`o$^{\ref{AFFIL::ItalyUandINFNBari}}$\orcid{0000-0002-9181-0345}}, 
  \mbox{N.~Randazzo$^{\ref{AFFIL::ItalyINFNCatania}}$}, 
  \mbox{R.~Rando$^{\ref{AFFIL::ItalyUPadovaandINFN}}$\orcid{0000-0001-6992-818X}}, 
  \mbox{T.~Ravel$^{\ref{AFFIL::FranceIRAPUToulouse}}$}, 
  \mbox{S.~Razzaque$^{\ref{AFFIL::SouthAfricaUJohannesburg},\ref{AFFIL::USAGWUWashingtonDC}}$\orcid{0000-0002-0130-2460}}, 
  \mbox{M.~Regeard$^{\ref{AFFIL::FranceAPCUParisCite}}$\orcid{0000-0002-3844-6003}}, 
  \mbox{P.~Reichherzer$^{\ref{AFFIL::UnitedKingdomUOxford},\ref{AFFIL::GermanyUBochum}}$\orcid{0000-0003-4513-8241}}, 
  \mbox{A.~Reimer$^{\ref{AFFIL::AustriaUInnsbruck}}$\orcid{0000-0001-8604-7077}}, 
  \mbox{O.~Reimer$^{\ref{AFFIL::AustriaUInnsbruck}}$\orcid{0000-0001-6953-1385}}, 
  \mbox{A.~Reisenegger$^{\ref{AFFIL::ChileUPontificiaCatolicadeChile},\ref{AFFIL::ChileUMCE}}$\orcid{0000-0003-4059-6796}}, 
  \mbox{T.~Reposeur$^{\ref{AFFIL::FranceLP2IUBordeaux}}$}, 
  \mbox{B.~Reville$^{\ref{AFFIL::GermanyMPIK}}$\orcid{0000-0002-3778-1432}}, 
  \mbox{W.~Rhode$^{\ref{AFFIL::GermanyUDortmundTU}}$\orcid{0000-0003-2636-5000}}, 
  \mbox{M.~Rib\'o$^{\ref{AFFIL::SpainICCUB}}$\orcid{0000-0002-9931-4557}}, 
  \mbox{T.~Richtler$^{\ref{AFFIL::ChileUdeConcepcion}}$}, 
  \mbox{F.~Rieger$^{\ref{AFFIL::GermanyMPIK}}$}, 
  \mbox{E.~Roache$^{\ref{AFFIL::USACfAHarvardSmithsonian}}$}, 
  \mbox{G.~Rodriguez~Fernandez$^{\ref{AFFIL::ItalyINFNRomaTorVergata}}$}, 
  \mbox{M.~D.~Rodr{\'\i}guez~Fr{\'\i}as$^{\ref{AFFIL::SpainUAlcala}}$\orcid{0000-0002-2550-4462}}, 
  \mbox{J.~J.~Rodr{\'\i}guez-V\'azquez$^{\ref{AFFIL::SpainCIEMAT}}$}, 
  \mbox{P.~Romano$^{\ref{AFFIL::ItalyOBrera}}$\orcid{0000-0003-0258-7469}}, 
  \mbox{G.~Romeo$^{\ref{AFFIL::ItalyOCatania}}$\orcid{0000-0003-3239-6057}}, 
  \mbox{J.~Rosado$^{\ref{AFFIL::SpainUCMAltasEnergias}}$}, 
  \mbox{G.~Rowell$^{\ref{AFFIL::AustraliaUAdelaide}}$\orcid{0000-0002-9516-1581}}, 
  \mbox{B.~Rudak$^{\ref{AFFIL::PolandNicolausCopernicusAstronomicalCenter}}$}, 
  \mbox{A.~J.~Ruiter$^{\ref{AFFIL::AustraliaUNewSouthWalesCanberra}}$\orcid{0000-0002-4794-6835}}, 
  \mbox{C.~B.~Rulten$^{\ref{AFFIL::UnitedKingdomUDurham}}$\orcid{0000-0001-7483-4348}}, 
  \mbox{F.~Russo$^{\ref{AFFIL::ItalyOASBologna}}$\orcid{0000-0002-3476-0839}}, 
  \mbox{I.~Sadeh$^{\ref{AFFIL::GermanyDESY}}$}, 
  \mbox{L.~Saha$^{\ref{AFFIL::USACfAHarvardSmithsonian}}$\orcid{0000-0002-3171-5039}}, 
  \mbox{T.~Saito$^{\ref{AFFIL::JapanUTokyoICRR}}$}, 
  \mbox{S.~Sakurai$^{\ref{AFFIL::JapanUTokyoICRR}}$}, 
  \mbox{H.~Salzmann$^{\ref{AFFIL::GermanyIAAT}}$}, 
  \mbox{D.~Sanchez$^{\ref{AFFIL::FranceLAPPUSavoieMontBlanc}}$}, 
  \mbox{M.~S\'anchez-Conde$^{\ref{AFFIL::SpainIFTUAMCSIC}}$\orcid{0000-0002-3849-9164}}, 
  \mbox{P.~Sangiorgi$^{\ref{AFFIL::ItalyIASFPalermo}}$\orcid{0000-0001-8138-9289}}, 
  \mbox{H.~Sano$^{\ref{AFFIL::JapanUTokyoICRR}}$\orcid{0000-0003-2062-5692}}, 
  \mbox{M.~Santander$^{\ref{AFFIL::USAUAlabamaTuscaloosa}}$\orcid{0000-0001-7297-8217}}, 
  \mbox{A.~Santangelo$^{\ref{AFFIL::GermanyIAAT}}$}, 
  \mbox{R.~Santos-Lima$^{\ref{AFFIL::BrazilIAGUSaoPaulo}}$\orcid{0000-0001-6880-4468}}, 
  \mbox{A.~Sanuy$^{\ref{AFFIL::SpainICCUB}}$}, 
  \mbox{T.~\v{S}ari\'c$^{\ref{AFFIL::CroatiaFESB}}$\orcid{0000-0001-8731-8369}}, 
  \mbox{A.~Sarkar$^{\ref{AFFIL::GermanyDESY}}$\orcid{0000-0002-7559-4339}}, 
  \mbox{S.~Sarkar$^{\ref{AFFIL::UnitedKingdomUOxford}}$\orcid{0000-0002-3542-858X}}, 
  \mbox{F.~G.~Saturni$^{\ref{AFFIL::ItalyORoma}}$\orcid{0000-0002-1946-7706}}, 
  \mbox{V.~Savchenko$^{\ref{AFFIL::SwitzerlandEPFLAstroObs}}$\orcid{0000-0001-6353-0808}}, 
  \mbox{A.~Scherer$^{\ref{AFFIL::ChileUPontificiaCatolicadeChile}}$}, 
  \mbox{P.~Schipani$^{\ref{AFFIL::ItalyOCapodimonte}}$\orcid{0000-0003-0197-589X}}, 
  \mbox{B.~Schleicher$^{\ref{AFFIL::GermanyUWurzburg},\ref{AFFIL::SwitzerlandETHZurich}}$}, 
  \mbox{P.~Schovanek$^{\ref{AFFIL::CzechRepublicFZU}}$}, 
  \mbox{J.~L.~Schubert$^{\ref{AFFIL::GermanyUDortmundTU}}$}, 
  \mbox{F.~Schussler$^{\ref{AFFIL::FranceCEAIRFUDPhP}}$\orcid{0000-0003-1500-6571}}, 
  \mbox{U.~Schwanke$^{\ref{AFFIL::GermanyUBerlin}}$\orcid{0000-0002-1229-278X}}, 
  \mbox{G.~Schwefer$^{\ref{AFFIL::GermanyMPIK}}$\orcid{0000-0002-2050-8413}}, 
  \mbox{S.~Scuderi$^{\ref{AFFIL::ItalyIASFMilano}}$\orcid{0000-0002-8637-2109}}, 
  \mbox{M.~Seglar~Arroyo$^{\ref{AFFIL::SpainIFAEBIST}}$\orcid{0000-0001-8654-409X}}, 
  \mbox{I.~Seitenzahl$^{\ref{AFFIL::AustraliaUNewSouthWalesCanberra}}$\orcid{0000-0002-5044-2988}}, 
  \mbox{O.~Sergijenko$^{\ref{AFFIL::UkraineAstObsofUKyiv},\ref{AFFIL::UkraineObsNASUkraine},\ref{AFFIL::PolandAGHCracowSTC}}$}, 
  \mbox{V.~Sguera$^{\ref{AFFIL::ItalyOASBologna}}$}, 
  \mbox{R.~Y.~Shang$^{\ref{AFFIL::USAUCLA}}$}, 
  \mbox{P.~Sharma$^{\ref{AFFIL::FranceIJCLab}}$}, 
  \mbox{G.~D.~S.~SIDIBE$^{\ref{AFFIL::FranceCEAIRFUDEDIP}}$}, 
  \mbox{L.~Sidoli$^{\ref{AFFIL::ItalyIASFMilano}}$\orcid{0000-0001-9705-2883}}, 
  \mbox{H.~Siejkowski$^{\ref{AFFIL::PolandCYFRONETAGH}}$\orcid{0000-0003-1673-2145}}, 
  \mbox{C.~Siqueira$^{\ref{AFFIL::BrazilIFSCUSaoPaulo}}$\orcid{0000-0001-5684-3849}}, 
  \mbox{P.~Sizun$^{\ref{AFFIL::FranceCEAIRFUDEDIP}}$\orcid{0000-0002-8895-3345}}, 
  \mbox{V.~Sliusar$^{\ref{AFFIL::SwitzerlandUGenevaISDC}}$\orcid{0000-0002-4387-9372}}, 
  \mbox{A.~Slowikowska$^{\ref{AFFIL::PolandTorunInstituteofAstronomy}}$\orcid{0000-0003-4525-3178}}, 
  \mbox{H.~Sol$^{\ref{AFFIL::FranceObservatoiredeParis}}$}, 
  \mbox{A.~Specovius$^{\ref{AFFIL::GermanyUErlangenECAP}}$\orcid{0000-0002-1156-4771}}, 
  \mbox{S.~T.~Spencer$^{\ref{AFFIL::GermanyUErlangenECAP},\ref{AFFIL::UnitedKingdomUOxford}}$\orcid{0000-0001-5516-1205}}, 
  \mbox{D.~Spiga$^{\ref{AFFIL::ItalyOBrera}}$\orcid{0000-0003-1163-7843}}, 
  \mbox{A.~Stamerra$^{\ref{AFFIL::ItalyORoma},\ref{AFFIL::ItalyCTAOBologna}}$\orcid{0000-0002-9430-5264}}, 
  \mbox{S.~Stani\v{c}$^{\ref{AFFIL::SloveniaUNovaGoricaCAC}}$\orcid{0000-0003-3344-8381}}, 
  \mbox{T.~Starecki$^{\ref{AFFIL::PolandWUTElectronics}}$\orcid{0000-0002-4730-6803}}, 
  \mbox{R.~Starling$^{\ref{AFFIL::UnitedKingdomULeicester}}$}, 
  \mbox{C.~Steppa$^{\ref{AFFIL::GermanyUPotsdam}}$}, 
  \mbox{T.~Stolarczyk$^{\ref{AFFIL::FranceCEAIRFUDAp}}$}, 
  \mbox{J.~Stri\v{s}kovi\'c$^{\ref{AFFIL::CroatiaUOsijek}}$}, 
  \mbox{M.~Strzys$^{\ref{AFFIL::JapanUTokyoICRR}}$\orcid{0000-0001-5049-1045}}, 
  \mbox{Y.~Suda$^{\ref{AFFIL::JapanUHiroshima}}$\orcid{0000-0002-2692-5891}}, 
  \mbox{T.~Suomij\"arvi$^{\ref{AFFIL::FranceIJCLab}}$\orcid{0000-0003-1422-258X}}, 
  \mbox{D.~Tak$^{\ref{AFFIL::GermanyDESY}}$\orcid{0000-0002-9852-2469}}, 
  \mbox{M.~Takahashi$^{\ref{AFFIL::JapanUNagoyaISEE}}$}, 
  \mbox{R.~Takeishi$^{\ref{AFFIL::JapanUTokyoICRR}}$\orcid{0000-0001-6335-5317}}, 
  \mbox{P.-H.~T.~Tam$^{\ref{AFFIL::JapanUTokyoICRR},\ref{AFFIL::ChinaUSunYatsen}}$\orcid{0000-0002-1262-7375}}, 
  \mbox{S.~J.~Tanaka$^{\ref{AFFIL::JapanUAoyamaGakuin}}$\orcid{0000-0002-8796-1992}}, 
  \mbox{T.~Tanaka$^{\ref{AFFIL::JapanUKonan}}$\orcid{0000-0002-4383-0368}}, 
  \mbox{K.~Terauchi$^{\ref{AFFIL::JapanUKyotoPhysicsandAstronomy}}$}, 
  \mbox{V.~Testa$^{\ref{AFFIL::ItalyORoma}}$\orcid{0000-0003-1033-1340}}, 
  \mbox{L.~Tibaldo$^{\ref{AFFIL::FranceIRAPUToulouse}}$\orcid{0000-0001-7523-570X}}, 
  \mbox{O.~Tibolla$^{\ref{AFFIL::UnitedKingdomUDurham}}$}, 
  \mbox{F.~Torradeflot$^{\ref{AFFIL::SpainPIC},\ref{AFFIL::SpainCIEMAT}}$\orcid{0000-0003-1160-1517}}, 
  \mbox{D.~F.~Torres$^{\ref{AFFIL::SpainICECSIC}}$}, 
  \mbox{E.~Torresi$^{\ref{AFFIL::ItalyOASBologna}}$\orcid{0000-0002-5201-010X}}, 
  \mbox{N.~Tothill$^{\ref{AFFIL::AustraliaUWesternSydney}}$\orcid{0000-0002-9931-5162}}, 
  \mbox{F.~Toussenel$^{\ref{AFFIL::FranceLPNHEUSorbonne}}$}, 
  \mbox{V.~Touzard$^{\ref{AFFIL::FranceIRAPUToulouse}}$}, 
  \mbox{A.~Tramacere$^{\ref{AFFIL::SwitzerlandUGenevaISDC}}$\orcid{0000-0002-8186-3793}}, 
  \mbox{P.~Travnicek$^{\ref{AFFIL::CzechRepublicFZU}}$}, 
  \mbox{G.~Tripodo$^{\ref{AFFIL::ItalyUPalermo},\ref{AFFIL::ItalyINFNCatania}}$}, 
  \mbox{S.~Truzzi$^{\ref{AFFIL::ItalyUSienaandINFN}}$}, 
  \mbox{A.~Tsiahina$^{\ref{AFFIL::FranceIRAPUToulouse}}$\orcid{0009-0006-6205-8728}}, 
  \mbox{A.~Tutone$^{\ref{AFFIL::ItalyIASFPalermo}}$}, 
  \mbox{M.~Vacula$^{\ref{AFFIL::CzechRepublicUOlomouc},\ref{AFFIL::CzechRepublicFZU}}$\orcid{0000-0003-4844-3962}}, 
  \mbox{B.~Vallage$^{\ref{AFFIL::FranceCEAIRFUDPhP}}$\orcid{0000-0003-1255-8506}}, 
  \mbox{P.~Vallania$^{\ref{AFFIL::ItalyINFNTorino},\ref{AFFIL::ItalyOTorino}}$}, 
  \mbox{R.~Vall\'es$^{\ref{AFFIL::SpainICECSIC}}$\orcid{0000-0001-7701-2163}}, 
  \mbox{C.~van~Eldik$^{\ref{AFFIL::GermanyUErlangenECAP}}$\orcid{0000-0001-9669-645X}}, 
  \mbox{J.~van~Scherpenberg$^{\ref{AFFIL::GermanyMPP}}$\orcid{0000-0002-6173-867X}}, 
  \mbox{J.~Vandenbroucke$^{\ref{AFFIL::USAUWisconsin}}$}, 
  \mbox{V.~Vassiliev$^{\ref{AFFIL::USAUCLA}}$}, 
  \mbox{P.~Venault$^{\ref{AFFIL::FranceCEAIRFUDEDIP}}$}, 
  \mbox{S.~Ventura$^{\ref{AFFIL::ItalyUSienaandINFN}}$}, 
  \mbox{S.~Vercellone$^{\ref{AFFIL::ItalyOBrera}}$\orcid{0000-0003-1163-1396}}, 
  \mbox{G.~Verna$^{\ref{AFFIL::ItalyUSienaandINFN}}$\orcid{0000-0001-5916-9028}}, 
  \mbox{A.~Viana$^{\ref{AFFIL::BrazilIFSCUSaoPaulo}}$}, 
  \mbox{N.~Viaux$^{\ref{AFFIL::ChileDepFisUTecnicaFedericoSantaMaria}}$}, 
  \mbox{A.~Vigliano$^{\ref{AFFIL::ItalyUUdineandINFNTrieste}}$}, 
  \mbox{J.~Vignatti$^{\ref{AFFIL::ChileUTecnicaFedericoSantaMaria}}$\orcid{0000-0002-1494-9562}}, 
  \mbox{C.~F.~Vigorito$^{\ref{AFFIL::ItalyINFNTorino},\ref{AFFIL::ItalyUTorino}}$\orcid{0000-0002-0069-9195}}, 
  \mbox{V.~Vitale$^{\ref{AFFIL::ItalyINFNRomaTorVergata}}$}, 
  \mbox{V.~Vodeb$^{\ref{AFFIL::SloveniaUNovaGoricaCAC}}$}, 
  \mbox{V.~Voisin$^{\ref{AFFIL::FranceLPNHEUSorbonne}}$}, 
  \mbox{S.~Vorobiov$^{\ref{AFFIL::SloveniaUNovaGoricaCAC}}$\orcid{0000-0001-8679-3424}}, 
  \mbox{G.~Voutsinas$^{\ref{AFFIL::SwitzerlandUGenevaDPNC}}$}, 
  \mbox{I.~Vovk$^{\ref{AFFIL::JapanUTokyoICRR}}$}, 
  \mbox{V.~Waegebaert$^{\ref{AFFIL::FranceIRAPUToulouse}}$}, 
  \mbox{S.~J.~Wagner$^{\ref{AFFIL::GermanyLSW}}$}, 
  \mbox{R.~Walter$^{\ref{AFFIL::SwitzerlandUGenevaISDC}}$}, 
  \mbox{M.~Ward$^{\ref{AFFIL::UnitedKingdomUDurham}}$}, 
  \mbox{M.~Wechakama$^{\ref{AFFIL::ThailandUKasetsart},\ref{AFFIL::ThailandNARIT}}$\orcid{0000-0001-8279-4550}}, 
  \mbox{R.~White$^{\ref{AFFIL::GermanyMPIK}}$}, 
  \mbox{A.~Wierzcholska$^{\ref{AFFIL::PolandIFJ}}$\orcid{0000-0003-4472-7204}}, 
  \mbox{M.~Will$^{\ref{AFFIL::GermanyMPP}}$\orcid{0000-0002-7504-2083}}, 
  \mbox{D.~A.~Williams$^{\ref{AFFIL::USASCIPP}}$\orcid{0000-0003-2740-9714}}, 
  \mbox{F.~Wohlleben$^{\ref{AFFIL::GermanyMPIK}}$\orcid{0000-0002-6451-4188}}, 
  \mbox{A.~Wolter$^{\ref{AFFIL::ItalyOBrera}}$\orcid{0000-0001-5840-9835}}, 
  \mbox{T.~Yamamoto$^{\ref{AFFIL::JapanUKonan}}$}, 
  \mbox{R.~Yamazaki$^{\ref{AFFIL::JapanUAoyamaGakuin}}$\orcid{0000-0002-1251-7889}}, 
  \mbox{L.~Yang$^{\ref{AFFIL::SouthAfricaUJohannesburg},\ref{AFFIL::ChinaUSunYatsen}}$}, 
  \mbox{T.~Yoshida$^{\ref{AFFIL::JapanUIbaraki}}$}, 
  \mbox{T.~Yoshikoshi$^{\ref{AFFIL::JapanUTokyoICRR}}$\orcid{0000-0002-6045-9839}}, 
  \mbox{M.~Zacharias$^{\ref{AFFIL::GermanyLSW},\ref{AFFIL::SouthAfricaNWU}}$\orcid{0000-0001-5801-3945}}, 
  \mbox{R.~Zanmar~Sanchez$^{\ref{AFFIL::ItalyOCatania}}$\orcid{0000-0002-6997-0887}}, 
  \mbox{D.~Zavrtanik$^{\ref{AFFIL::SloveniaUNovaGoricaCAC}}$\orcid{0000-0002-4596-1521}}, 
  \mbox{M.~Zavrtanik$^{\ref{AFFIL::SloveniaUNovaGoricaCAC}}$}, 
  \mbox{A.~A.~Zdziarski$^{\ref{AFFIL::PolandNicolausCopernicusAstronomicalCenter}}$}, 
  \mbox{A.~Zech$^{\ref{AFFIL::FranceObservatoiredeParis}}$\orcid{0000-0002-4388-5625}}, 
  \mbox{V.~I.~Zhdanov$^{\ref{AFFIL::UkraineAstObsofUKyiv}}$\orcid{0000-0003-3690-483X}}, 
  \mbox{K.~Zi\k{e}tara$^{\ref{AFFIL::PolandUJagiellonian}}$}, 
  \mbox{M.~\v{Z}ivec$^{\ref{AFFIL::SloveniaUNovaGoricaCAC}}$}, 
  \mbox{J.~Zuriaga-Puig$^{\ref{AFFIL::SpainIFTUAMCSIC}}$\orcid{0000-0003-0652-6700}}
\twocolumn
\section*{Affiliations}
\begin{enumerate}[label=$^{\arabic*}$,ref=\arabic*,leftmargin=1.5em,labelsep=0.25em,labelwidth=1.25em]
\item Department of Physics, Tokai University, 4-1-1, Kita-Kaname, Hiratsuka, Kanagawa 259-1292, Japan\label{AFFIL::JapanUTokai}
\item Institute for Cosmic Ray Research, University of Tokyo, 5-1-5, Kashiwa-no-ha, Kashiwa, Chiba 277-8582, Japan\label{AFFIL::JapanUTokyoICRR}
\item University of Alabama, Tuscaloosa, Department of Physics and Astronomy, Gallalee Hall, Box 870324 Tuscaloosa, AL 35487-0324, USA\label{AFFIL::USAUAlabamaTuscaloosa}
\item Universit\'e C\^ote d'Azur, Observatoire de la C\^ote d'Azur, CNRS, Laboratoire Lagrange, France\label{AFFIL::FranceOCotedAzur}
\item Laboratoire Leprince-Ringuet, CNRS/IN2P3, \'Ecole polytechnique, Institut Polytechnique de Paris, 91120 Palaiseau, France\label{AFFIL::FranceLLREcolePolytechnique}
\item Departament de F{\'\i}sica Qu\`antica i Astrof{\'\i}sica, Institut de Ci\`encies del Cosmos, Universitat de Barcelona, IEEC-UB, Mart{\'\i} i Franqu\`es, 1, 08028, Barcelona, Spain\label{AFFIL::SpainICCUB}
\item Instituto de Astrof{\'\i}sica de Andaluc{\'\i}a-CSIC, Glorieta de la Astronom{\'\i}a s/n, 18008, Granada, Spain\label{AFFIL::SpainIAACSIC}
\item Pontificia Universidad Cat\'olica de Chile, Av. Libertador Bernardo O'Higgins 340, Santiago, Chile\label{AFFIL::ChileUPontificiaCatolicadeChile}
\item IPARCOS-UCM, Instituto de F{\'\i}sica de Part{\'\i}culas y del Cosmos, and EMFTEL Department, Universidad Complutense de Madrid, E-28040 Madrid, Spain\label{AFFIL::SpainUCMAltasEnergias}
\item Instituto de F{\'\i}sica Te\'orica UAM/CSIC and Departamento de F{\'\i}sica Te\'orica, Universidad Aut\'onoma de Madrid, c/ Nicol\'as Cabrera 13-15, Campus de Cantoblanco UAM, 28049 Madrid, Spain\label{AFFIL::SpainIFTUAMCSIC}
\item LUTH, GEPI and LERMA, Observatoire de Paris, Universit\'e PSL, Universit\'e Paris Cit\'e, CNRS, 5 place Jules Janssen, 92190, Meudon, France\label{AFFIL::FranceObservatoiredeParis}
\item INAF - Osservatorio Astrofisico di Arcetri, Largo E. Fermi, 5 - 50125 Firenze, Italy\label{AFFIL::ItalyOArcetri}
\item INAF - Osservatorio Astronomico di Roma, Via di Frascati 33, 00040, Monteporzio Catone, Italy\label{AFFIL::ItalyORoma}
\item T\"UB\.ITAK Research Institute for Fundamental Sciences, 41470 Gebze, Kocaeli, Turkey\label{AFFIL::TurkeyTubitak}
\item INFN Sezione di Napoli, Via Cintia, ed. G, 80126 Napoli, Italy\label{AFFIL::ItalyINFNNapoli}
\item INFN Sezione di Padova, Via Marzolo 8, 35131 Padova, Italy\label{AFFIL::ItalyINFNPadova}
\item Laboratoire Univers et Particules de Montpellier, Universit\'e de Montpellier, CNRS/IN2P3, CC 72, Place Eug\`ene Bataillon, F-34095 Montpellier Cedex 5, France\label{AFFIL::FranceLUPMUMontpellier}
\item Kapteyn Astronomical Institute, University of Groningen, Landleven 12, 9747 AD, Groningen, The Netherlands\label{AFFIL::NetherlandsUGroningen}
\item Instituto de F{\'\i}sica de S\~ao Carlos, Universidade de S\~ao Paulo, Av. Trabalhador S\~ao-carlense, 400 - CEP 13566-590, S\~ao Carlos, SP, Brazil\label{AFFIL::BrazilIFSCUSaoPaulo}
\item Astroparticle Physics, Department of Physics, TU Dortmund University, Otto-Hahn-Str. 4a, 44227 Dortmund, Germany\label{AFFIL::GermanyUDortmundTU}
\item Department of Physics, Chemistry \& Material Science, University of Namibia, Private Bag 13301, Windhoek, Namibia\label{AFFIL::NamibiaUNamibia}
\item Centre for Space Research, North-West University, Potchefstroom, 2520, South Africa\label{AFFIL::SouthAfricaNWU}
\item School of Physics and Astronomy, Monash University, Melbourne, Victoria 3800, Australia\label{AFFIL::AustraliaUMonash}
\item Department of Astronomy, University of Geneva, Chemin d'Ecogia 16, CH-1290 Versoix, Switzerland\label{AFFIL::SwitzerlandUGenevaISDC}
\item Faculty of Science and Technology, Universidad del Azuay, Cuenca, Ecuador.\label{AFFIL::EcuadorUAzuay}
\item Deutsches Elektronen-Synchrotron, Platanenallee 6, 15738 Zeuthen, Germany\label{AFFIL::GermanyDESY}
\item Centro Brasileiro de Pesquisas F{\'\i}sicas, Rua Xavier Sigaud 150, RJ 22290-180, Rio de Janeiro, Brazil\label{AFFIL::BrazilCBPF}
\item Instituto de Astronomia, Geof{\'\i}sica e Ci\^encias Atmosf\'ericas - Universidade de S\~ao Paulo, Cidade Universit\'aria, R. do Mat\~ao, 1226, CEP 05508-090, S\~ao Paulo, SP, Brazil\label{AFFIL::BrazilIAGUSaoPaulo}
\item INFN Sezione di Padova and Universit\`a degli Studi di Padova, Via Marzolo 8, 35131 Padova, Italy\label{AFFIL::ItalyUPadovaandINFN}
\item Institut f\"ur Physik \& Astronomie, Universit\"at Potsdam, Karl-Liebknecht-Strasse 24/25, 14476 Potsdam, Germany\label{AFFIL::GermanyUPotsdam}
\item University of the Witwatersrand, 1 Jan Smuts Avenue, Braamfontein, 2000 Johannesburg, South Africa\label{AFFIL::SouthAfricaUWitwatersrand}
\item Institut f\"ur Theoretische Physik, Lehrstuhl IV: Plasma-Astroteilchenphysik, Ruhr-Universit\"at Bochum, Universit\"atsstra{\ss}e 150, 44801 Bochum, Germany\label{AFFIL::GermanyUBochum}
\item INFN Sezione di Roma Tor Vergata, Via della Ricerca Scientifica 1, 00133 Rome, Italy\label{AFFIL::ItalyINFNRomaTorVergata}
\item Center for Astrophysics | Harvard \& Smithsonian, 60 Garden St, Cambridge, MA 02138, USA\label{AFFIL::USACfAHarvardSmithsonian}
\item CIEMAT, Avda. Complutense 40, 28040 Madrid, Spain\label{AFFIL::SpainCIEMAT}
\item Max-Planck-Institut f\"ur Kernphysik, Saupfercheckweg 1, 69117 Heidelberg, Germany\label{AFFIL::GermanyMPIK}
\item Max-Planck-Institut f\"ur Physik, F\"ohringer Ring 6, 80805 M\"unchen, Germany\label{AFFIL::GermanyMPP}
\item Pidstryhach Institute for Applied Problems in Mechanics and Mathematics NASU, 3B Naukova Street, Lviv, 79060, Ukraine\label{AFFIL::UkraineIAPMMLviv}
\item Univ. Savoie Mont Blanc, CNRS, Laboratoire d'Annecy de Physique des Particules - IN2P3, 74000 Annecy, France\label{AFFIL::FranceLAPPUSavoieMontBlanc}
\item Center for Astrophysics and Cosmology (CAC), University of Nova Gorica, Nova Gorica, Slovenia\label{AFFIL::SloveniaUNovaGoricaCAC}
\item Institut f\"ur Astronomie und Astrophysik, Universit\"at T\"ubingen, Sand 1, 72076 T\"ubingen, Germany\label{AFFIL::GermanyIAAT}
\item ETH Z\"urich, Institute for Particle Physics and Astrophysics, Otto-Stern-Weg 5, 8093 Z\"urich, Switzerland\label{AFFIL::SwitzerlandETHZurich}
\item Politecnico di Bari, via Orabona 4, 70124 Bari, Italy\label{AFFIL::ItalyPolitecnicoBari}
\item INFN Sezione di Bari, via Orabona 4, 70126 Bari, Italy\label{AFFIL::ItalyINFNBari}
\item Institut de Fisica d'Altes Energies (IFAE), The Barcelona Institute of Science and Technology, Campus UAB, 08193 Bellaterra (Barcelona), Spain\label{AFFIL::SpainIFAEBIST}
\item FZU - Institute of Physics of the Czech Academy of Sciences, Na Slovance 1999/2, 182 21 Praha 8, Czech Republic\label{AFFIL::CzechRepublicFZU}
\item Sorbonne Universit\'e, CNRS/IN2P3, Laboratoire de Physique Nucl\'eaire et de Hautes Energies, LPNHE, 4 place Jussieu, 75005 Paris, France\label{AFFIL::FranceLPNHEUSorbonne}
\item INAF - Osservatorio Astronomico di Brera, Via Brera 28, 20121 Milano, Italy\label{AFFIL::ItalyOBrera}
\item INFN Sezione di Pisa, Edificio C {\textendash} Polo Fibonacci, Largo Bruno Pontecorvo 3, 56127 Pisa\label{AFFIL::ItalyINFNPisa}
\item University of Zagreb, Faculty of electrical engineering and computing, Unska 3, 10000 Zagreb, Croatia\label{AFFIL::CroatiaUZagreb}
\item IRFU, CEA, Universit\'e Paris-Saclay, B\^at 141, 91191 Gif-sur-Yvette, France\label{AFFIL::FranceCEAIRFUDPhP}
\item School of Physics, Chemistry and Earth Sciences, University of Adelaide, Adelaide SA 5005, Australia\label{AFFIL::AustraliaUAdelaide}
\item INAF - Osservatorio di Astrofisica e Scienza dello spazio di Bologna, Via Piero Gobetti 93/3, 40129  Bologna, Italy\label{AFFIL::ItalyOASBologna}
\item Dublin Institute for Advanced Studies, 31 Fitzwilliam Place, Dublin 2, Ireland\label{AFFIL::IrelandDIAS}
\item Centre for Advanced Instrumentation, Department of Physics, Durham University, South Road, Durham, DH1 3LE, United Kingdom\label{AFFIL::UnitedKingdomUDurham}
\item INFN Sezione di Trieste and Universit\`a degli Studi di Udine, Via delle Scienze 208, 33100 Udine, Italy\label{AFFIL::ItalyUUdineandINFNTrieste}
\item University of Geneva - D\'epartement de physique nucl\'eaire et corpusculaire, 24 rue du G\'en\'eral-Dufour, 1211 Gen\`eve 4, Switzerland\label{AFFIL::SwitzerlandUGenevaDPNC}
\item Armagh Observatory and Planetarium, College Hill, Armagh BT61 9DG, United Kingdom\label{AFFIL::UnitedKingdomArmaghObservatoryandPlanetarium}
\item School of Physics, University of New South Wales, Sydney NSW 2052, Australia\label{AFFIL::AustraliaUNewSouthWales}
\item Universit\'e Paris-Saclay, Universit\'e Paris Cit\'e, CEA, CNRS, AIM, F-91191 Gif-sur-Yvette Cedex, France\label{AFFIL::FranceCEAIRFUDAp}
\item Cherenkov Telescope Array Observatory, Saupfercheckweg 1, 69117 Heidelberg, Germany\label{AFFIL::GermanyCTAOHeidelberg}
\item Unitat de F{\'\i}sica de les Radiacions, Departament de F{\'\i}sica, and CERES-IEEC, Universitat Aut\`onoma de Barcelona, Edifici C3, Campus UAB, 08193 Bellaterra, Spain\label{AFFIL::SpainUABandCERESIEEC}
\item Department of Physics, Faculty of Science, Kasetsart University, 50 Ngam Wong Wan Rd., Lat Yao, Chatuchak, Bangkok, 10900, Thailand\label{AFFIL::ThailandUKasetsart}
\item National Astronomical Research Institute of Thailand, 191 Huay Kaew Rd., Suthep, Muang, Chiang Mai, 50200, Thailand\label{AFFIL::ThailandNARIT}
\item INAF - Istituto di Astrofisica Spaziale e Fisica Cosmica di Palermo, Via U. La Malfa 153, 90146 Palermo, Italy\label{AFFIL::ItalyIASFPalermo}
\item Universidade Cruzeiro do Sul, N\'ucleo de Astrof{\'\i}sica Te\'orica (NAT/UCS), Rua Galv\~ao Bueno 8687, Bloco B, sala 16, Libertade 01506-000 - S\~ao Paulo, Brazil\label{AFFIL::BrazilUCruzeirodoSul}
\item Lund Observatory, Lund University, Box 43, SE-22100 Lund, Sweden\label{AFFIL::SwedenLundObservatory}
\item Aix Marseille Univ, CNRS/IN2P3, CPPM, Marseille, France\label{AFFIL::FranceCPPMUAixMarseille}
\item INAF - Osservatorio Astronomico di Capodimonte, Via Salita Moiariello 16, 80131 Napoli, Italy\label{AFFIL::ItalyOCapodimonte}
\item Universidad de Alcal\'a - Space \& Astroparticle group, Facultad de Ciencias, Campus Universitario Ctra. Madrid-Barcelona, Km. 33.600 28871 Alcal\'a de Henares (Madrid), Spain\label{AFFIL::SpainUAlcala}
\item Escola de Engenharia de Lorena, Universidade de S\~ao Paulo, \'Area I - Estrada Municipal do Campinho, s/n{\textdegree}, CEP 12602-810, Pte. Nova, Lorena, Brazil\label{AFFIL::BrazilEELUSaoPaulo}
\item INFN Sezione di Bari and Universit\`a degli Studi di Bari, via Orabona 4, 70124 Bari, Italy\label{AFFIL::ItalyUandINFNBari}
\item Universit\'e Paris Cit\'e, CNRS, Astroparticule et Cosmologie, F-75013 Paris, France\label{AFFIL::FranceAPCUParisCite}
\item Dublin City University, Glasnevin, Dublin 9, Ireland\label{AFFIL::IrelandDCU}
\item INFN Sezione di Torino, Via P. Giuria 1, 10125 Torino, Italy\label{AFFIL::ItalyINFNTorino}
\item Dipartimento di Fisica - Universit\`a degli Studi di Torino, Via Pietro Giuria 1 - 10125 Torino, Italy\label{AFFIL::ItalyUTorino}
\item Universidade Federal Do Paran\'a - Setor Palotina, Departamento de Engenharias e Exatas, Rua Pioneiro, 2153, Jardim Dallas, CEP: 85950-000 Palotina, Paran\'a, Brazil\label{AFFIL::BrazilUFPR}
\item INAF - Osservatorio Astrofisico di Catania, Via S. Sofia, 78, 95123 Catania, Italy\label{AFFIL::ItalyOCatania}
\item Universidad de Valpara{\'\i}so, Blanco 951, Valparaiso, Chile\label{AFFIL::ChileUdeValparaiso}
\item University of Wisconsin, Madison, 500 Lincoln Drive, Madison, WI, 53706, USA\label{AFFIL::USAUWisconsin}
\item Department of Physics and Technology, University of Bergen, Museplass 1, 5007 Bergen, Norway\label{AFFIL::NorwayUBergen}
\item INAF - Istituto di Radioastronomia, Via Gobetti 101, 40129 Bologna, Italy\label{AFFIL::ItalyRadioastronomiaINAF}
\item INAF - Istituto Nazionale di Astrofisica, Viale del Parco Mellini 84, 00136 Rome, Italy\label{AFFIL::ItalyINAF}
\item IRFU/DEDIP, CEA, Universit\'e Paris-Saclay, Bat 141, 91191 Gif-sur-Yvette, France\label{AFFIL::FranceCEAIRFUDEDIP}
\item Universit\'a degli Studi di Napoli {\textquotedblleft}Federico II{\textquotedblright} - Dipartimento di Fisica {\textquotedblleft}E. Pancini{\textquotedblright}, Complesso universitario di Monte Sant'Angelo, Via Cintia - 80126 Napoli, Italy\label{AFFIL::ItalyUNapoli}
\item CCTVal, Universidad T\'ecnica Federico Santa Mar{\'\i}a, Avenida Espa\~na 1680, Valpara{\'\i}so, Chile\label{AFFIL::ChileUTecnicaFedericoSantaMaria}
\item Institute for Theoretical Physics and Astrophysics, Universit\"at W\"urzburg, Campus Hubland Nord, Emil-Fischer-Str. 31, 97074 W\"urzburg, Germany\label{AFFIL::GermanyUWurzburg}
\item Friedrich-Alexander-Universit\"at Erlangen-N\"urnberg, Erlangen Centre for Astroparticle Physics, Nikolaus-Fiebiger-Str. 2, 91058 Erlangen, Germany\label{AFFIL::GermanyUErlangenECAP}
\item Universit\'e Paris-Saclay, CNRS/IN2P3, IJCLab, 91405 Orsay, France\label{AFFIL::FranceIJCLab}
\item Department of Astronomy and Astrophysics, University of Chicago, 5640 S Ellis Ave, Chicago, Illinois, 60637, USA\label{AFFIL::USAUChicagoDAA}
\item LAPTh, CNRS, USMB, F-74940 Annecy, France\label{AFFIL::FranceLAPTh}
\item Santa Cruz Institute for Particle Physics and Department of Physics, University of California, Santa Cruz, 1156 High Street, Santa Cruz, CA 95064, USA\label{AFFIL::USASCIPP}
\item University School for Advanced Studies IUSS Pavia, Palazzo del Broletto, Piazza della Vittoria 15, 27100 Pavia, Italy\label{AFFIL::ItalyIUSSPaviaINAF}
\item INAF - Istituto di Astrofisica Spaziale e Fisica Cosmica di Milano, Via A. Corti 12, 20133 Milano, Italy\label{AFFIL::ItalyIASFMilano}
\item Escola de Artes, Ci\^encias e Humanidades, Universidade de S\~ao Paulo, Rua Arlindo Bettio, CEP 03828-000, 1000 S\~ao Paulo, Brazil\label{AFFIL::BrazilEACHUSaoPaulo}
\item Astronomical Observatory of Taras Shevchenko National University of Kyiv, 3 Observatorna Street, Kyiv, 04053, Ukraine\label{AFFIL::UkraineAstObsofUKyiv}
\item The University of Manitoba, Dept of Physics and Astronomy, Winnipeg, Manitoba R3T 2N2, Canada\label{AFFIL::CanadaUManitoba}
\item RIKEN, Institute of Physical and Chemical Research, 2-1 Hirosawa, Wako, Saitama, 351-0198, Japan\label{AFFIL::JapanRIKEN}
\item INFN Sezione di Roma La Sapienza, P.le Aldo Moro, 2 - 00185 Roma, Italy\label{AFFIL::ItalyINFNRomaLaSapienza}
\item INFN Sezione di Perugia and Universit\`a degli Studi di Perugia, Via A. Pascoli, 06123 Perugia, Italy\label{AFFIL::ItalyUPerugiaandINFN}
\item INAF - Istituto di Astrofisica e Planetologia Spaziali (IAPS), Via del Fosso del Cavaliere 100, 00133 Roma, Italy\label{AFFIL::ItalyIAPS}
\item Department of Physics, Nagoya University, Chikusa-ku, Nagoya, 464-8602, Japan\label{AFFIL::JapanUNagoya}
\item Alikhanyan National Science Laboratory, Yerevan Physics Institute, 2 Alikhanyan Brothers St., 0036, Yerevan, Armenia\label{AFFIL::ArmeniaNSLAlikhanyan}
\item INFN Sezione di Catania, Via S. Sofia 64, 95123 Catania, Italy\label{AFFIL::ItalyINFNCatania}
\item Universit\'e Paris Cit\'e, CNRS, CEA, Astroparticule et Cosmologie, F-75013 Paris, France\label{AFFIL::FranceAPCUParisCiteCEAaffiliatedpersonnel}
\item Universidad Andres Bello, Rep\'ublica 252, Santiago, Chile\label{AFFIL::ChileUAndresBello}
\item Universidad Nacional Aut\'onoma de M\'exico, Delegaci\'on Coyoac\'an, 04510 Ciudad de M\'exico, Mexico\label{AFFIL::MexicoUNAMMexico}
\item N\'ucleo de Astrof{\'\i}sica e Cosmologia (Cosmo-ufes) \& Departamento de F{\'\i}sica, Universidade Federal do Esp{\'\i}rito Santo (UFES), Av. Fernando Ferrari, 514. 29065-910. Vit\'oria-ES, Brazil\label{AFFIL::BrazilUFES}
\item Astrophysics Research Center of the Open University (ARCO), The Open University of Israel, P.O. Box 808, Ra{\textquoteright}anana 4353701, Israel\label{AFFIL::IsraelOpenUniversityofIsrael}
\item Department of Physics, The George Washington University, Washington, DC 20052, USA\label{AFFIL::USAGWUWashingtonDC}
\item University of Liverpool, Oliver Lodge Laboratory, Liverpool L69 7ZE, United Kingdom\label{AFFIL::UnitedKingdomULiverpool}
\item King's College London, Strand, London, WC2R 2LS, United Kingdom\label{AFFIL::UnitedKingdomKingsCollege}
\item Department of Physics, Yamagata University, Yamagata, Yamagata 990-8560, Japan\label{AFFIL::JapanUYamagata}
\item Learning and Education Development Center, Yamanashi-Gakuin University, Kofu, Yamanashi 400-8575, Japan\label{AFFIL::JapanUYamanashiGakuin}
\item IRAP, Universit\'e de Toulouse, CNRS, CNES, UPS, 9 avenue Colonel Roche, 31028 Toulouse, Cedex 4, France\label{AFFIL::FranceIRAPUToulouse}
\item Universit\"at Innsbruck, Institut f\"ur Astro- und Teilchenphysik, Technikerstr. 25/8, 6020 Innsbruck, Austria\label{AFFIL::AustriaUInnsbruck}
\item Palack\'y University Olomouc, Faculty of Science, Joint Laboratory of Optics of Palack\'y University and Institute of Physics of the Czech Academy of Sciences, 17. listopadu 1192/12, 779 00 Olomouc, Czech Republic\label{AFFIL::CzechRepublicUOlomouc}
\item Finnish Centre for Astronomy with ESO, University of Turku, Finland, FI-20014 University of Turku, Finland\label{AFFIL::FinlandUTurku}
\item Josip Juraj Strossmayer University of Osijek, Trg Ljudevita Gaja 6, 31000 Osijek, Croatia\label{AFFIL::CroatiaUOsijek}
\item Gran Sasso Science Institute (GSSI), Viale Francesco Crispi 7, 67100 L{\textquoteright}Aquila, Italy and INFN-Laboratori Nazionali del Gran Sasso (LNGS), via G. Acitelli 22, 67100 Assergi (AQ), Italy\label{AFFIL::ItalyGSSIandINFNAquila}
\item Dipartimento di Scienze Fisiche e Chimiche, Universit\`a degli Studi dell'Aquila and GSGC-LNGS-INFN, Via Vetoio 1, L'Aquila, 67100, Italy\label{AFFIL::ItalyUandINFNAquila}
\item Faculty of Physics and Applied Computer Science,  University of L\'od\'z, ul. Pomorska 149-153, 90-236 L\'od\'z, Poland\label{AFFIL::PolandULodz}
\item Astronomical Observatory, Jagiellonian University, ul. Orla 171, 30-244 Cracow, Poland\label{AFFIL::PolandUJagiellonian}
\item Landessternwarte, Zentrum f\"ur Astronomie  der Universit\"at Heidelberg, K\"onigstuhl 12, 69117 Heidelberg, Germany\label{AFFIL::GermanyLSW}
\item Univ. Grenoble Alpes, CNRS, IPAG, 414 rue de la Piscine, Domaine Universitaire, 38041 Grenoble Cedex 9, France\label{AFFIL::FranceIPAGUGrenobleAlpes}
\item Astronomical Institute of the Czech Academy of Sciences, Bocni II 1401 - 14100 Prague, Czech Republic\label{AFFIL::CzechRepublicASU}
\item Department of Physics and Astronomy, University of Utah, Salt Lake City, UT 84112-0830, USA\label{AFFIL::USAUUtah}
\item Nicolaus Copernicus Astronomical Center, Polish Academy of Sciences, ul. Bartycka 18, 00-716 Warsaw, Poland\label{AFFIL::PolandNicolausCopernicusAstronomicalCenter}
\item Institute of Particle and Nuclear Studies,  KEK (High Energy Accelerator Research Organization), 1-1 Oho, Tsukuba, 305-0801, Japan\label{AFFIL::JapanKEK}
\item School of Physics and Astronomy, University of Leicester, Leicester, LE1 7RH, United Kingdom\label{AFFIL::UnitedKingdomULeicester}
\item Western Sydney University, Locked Bag 1797, Penrith, NSW 2751, Australia\label{AFFIL::AustraliaUWesternSydney}
\item Universit\'e Bordeaux, CNRS, LP2I Bordeaux, UMR 5797, 19 Chemin du Solarium, F-33170 Gradignan, France\label{AFFIL::FranceLP2IUBordeaux}
\item INFN Sezione di Trieste and Universit\`a degli Studi di Trieste, Via Valerio 2 I, 34127 Trieste, Italy\label{AFFIL::ItalyUandINFNTrieste}
\item Instituto de Astrof{\'\i}sica de Canarias and Departamento de Astrof{\'\i}sica, Universidad de La Laguna, La Laguna, Tenerife, Spain\label{AFFIL::SpainIAC}
\item Escuela Polit\'ecnica Superior de Ja\'en, Universidad de Ja\'en, Campus Las Lagunillas s/n, Edif. A3, 23071 Ja\'en, Spain\label{AFFIL::SpainUJaen}
\item Anton Pannekoek Institute/GRAPPA, University of Amsterdam, Science Park 904 1098 XH Amsterdam, The Netherlands\label{AFFIL::NetherlandsUAmsterdam}
\item Saha Institute of Nuclear Physics, A CI of Homi Bhabha National Institute, Kolkata 700064, West Bengal, India\label{AFFIL::IndiaSahaInstitute}
\item Universit\`a degli studi di Catania, Dipartimento di Fisica e Astronomia {\textquotedblleft}Ettore Majorana{\textquotedblright}, Via S. Sofia 64, 95123 Catania, Italy\label{AFFIL::ItalyUCatania}
\item Dipartimento di Fisica e Chimica {\textquotedblleft}E. Segr\`e{\textquotedblright}, Universit\`a degli Studi di Palermo, Via Archirafi 36, 90123, Palermo, Italy\label{AFFIL::ItalyUPalermo}
\item UCM-ELEC group, EMFTEL Department, University Complutense of Madrid, 28040 Madrid, Spain\label{AFFIL::SpainUCMElectronica}
\item Departamento de Ingenier{\'\i}a El\'ectrica, Universidad Pontificia de Comillas - ICAI, 28015 Madrid\label{AFFIL::SpainUPCMadrid}
\item Universidad de Chile, Av. Libertador Bernardo O'Higgins 1058, Santiago, Chile\label{AFFIL::ChileUdeChile}
\item Institute of Space Sciences (ICE, CSIC), and Institut d'Estudis Espacials de Catalunya (IEEC), and Instituci\'o Catalana de Recerca I Estudis Avan\c{c}ats (ICREA), Campus UAB, Carrer de Can Magrans, s/n 08193 Cerdanyola del Vall\'es, Spain\label{AFFIL::SpainICECSIC}
\item The Henryk Niewodnicza\'nski Institute of Nuclear Physics, Polish Academy of Sciences, ul. Radzikowskiego 152, 31-342 Cracow, Poland\label{AFFIL::PolandIFJ}
\item IPARCOS Institute, Faculty of Physics (UCM), 28040 Madrid, Spain\label{AFFIL::SpainIPARCOSInstitute}
\item Department of Physics, Konan University, Kobe, Hyogo, 658-8501, Japan\label{AFFIL::JapanUKonan}
\item Hiroshima Astrophysical Science Center, Hiroshima University, Higashi-Hiroshima, Hiroshima 739-8526, Japan\label{AFFIL::JapanHASC}
\item Department of Physics, Columbia University, 538 West 120th Street, New York, NY 10027, USA\label{AFFIL::USABarnardCollegeColumbiaUniversity}
\item School of Allied Health Sciences, Kitasato University, Sagamihara, Kanagawa 228-8555, Japan\label{AFFIL::JapanUKitasato}
\item Kavli Institute for Particle Astrophysics and Cosmology, Stanford University, Stanford, CA 94305, USA\label{AFFIL::USAStanford}
\item University of Bia{\l}ystok, Faculty of Physics, ul. K. Cio{\l}kowskiego 1L, 15-245 Bia{\l}ystok, Poland\label{AFFIL::PolandUBiaystok}
\item Charles University, Institute of Particle \& Nuclear Physics, V Hole\v{s}ovi\v{c}k\'ach 2, 180 00 Prague 8, Czech Republic\label{AFFIL::CzechRepublicUPrague}
\item Astronomical Observatory of Ivan Franko National University of Lviv, 8 Kyryla i Mephodia Street, Lviv, 79005, Ukraine\label{AFFIL::UkraineAstObsofULviv}
\item Institute for Space{\textemdash}Earth Environmental Research, Nagoya University, Furo-cho, Chikusa-ku, Nagoya 464-8601, Japan\label{AFFIL::JapanUNagoyaISEE}
\item Kobayashi{\textemdash}Maskawa Institute for the Origin of Particles and the Universe, Nagoya University, Furo-cho, Chikusa-ku, Nagoya 464-8602, Japan\label{AFFIL::JapanUNagoyaKMI}
\item INAF - Osservatorio Astronomico di Palermo {\textquotedblleft}G.S. Vaiana{\textquotedblright}, Piazza del Parlamento 1, 90134 Palermo, Italy\label{AFFIL::ItalyOPalermo}
\item Department of Physics and Astronomy, University of California, Los Angeles, CA 90095, USA\label{AFFIL::USAUCLA}
\item Graduate School of Technology, Industrial and Social Sciences, Tokushima University, Tokushima 770-8506, Japan\label{AFFIL::JapanUTokushima}
\item School of Physics \& Center for Relativistic Astrophysics, Georgia Institute of Technology, 837 State Street, Atlanta, Georgia, 30332-0430, USA\label{AFFIL::USAGeorgiaTech}
\item University of Pisa, Largo B. Pontecorvo 3, 56127 Pisa, Italy \label{AFFIL::ItalyUPisa}
\item University of Rijeka, Faculty of Physics, Radmile Matejcic 2, 51000 Rijeka, Croatia\label{AFFIL::CroatiaURijeka}
\item Rudjer Boskovic Institute, Bijenicka 54, 10 000 Zagreb, Croatia\label{AFFIL::CroatiaIRB}
\item INAF - Osservatorio Astronomico di Padova, Vicolo dell'Osservatorio 5, 35122 Padova, Italy\label{AFFIL::ItalyOPadova}
\item INAF - Osservatorio Astronomico di Padova and INFN Sezione di Trieste, gr. coll. Udine, Via delle Scienze 208 I-33100 Udine, Italy\label{AFFIL::ItalyOandINFNTrieste}
\item INFN and Universit\`a degli Studi di Siena, Dipartimento di Scienze Fisiche, della Terra e dell'Ambiente (DSFTA), Sezione di Fisica, Via Roma 56, 53100 Siena, Italy\label{AFFIL::ItalyUSienaandINFN}
\item Centre for Astro-Particle Physics (CAPP) and Department of Physics, University of Johannesburg, PO Box 524, Auckland Park 2006, South Africa\label{AFFIL::SouthAfricaUJohannesburg}
\item University of Oxford, Department of Physics, Clarendon Laboratory, Parks Road, Oxford, OX1 3PU, United Kingdom\label{AFFIL::UnitedKingdomUOxford}
\item Departamento de F{\'\i}sica, Facultad de Ciencias B\'asicas, Universidad Metropolitana de Ciencias de la Educaci\'on, Avenida Jos\'e Pedro Alessandri 774, \~Nu\~noa, Santiago, Chile\label{AFFIL::ChileUMCE}
\item Departamento de Astronom{\'\i}a, Universidad de Concepci\'on, Barrio Universitario S/N, Concepci\'on, Chile\label{AFFIL::ChileUdeConcepcion}
\item University of New South Wales, School of Science, Australian Defence Force Academy, Canberra, ACT 2600, Australia \label{AFFIL::AustraliaUNewSouthWalesCanberra}
\item University of Split  - FESB, R. Boskovica 32, 21 000 Split, Croatia\label{AFFIL::CroatiaFESB}
\item EPFL Laboratoire d{\textquoteright}astrophysique, Observatoire de Sauverny, CH-1290 Versoix, Switzerland\label{AFFIL::SwitzerlandEPFLAstroObs}
\item Department of Physics, Humboldt University Berlin, Newtonstr. 15, 12489 Berlin, Germany\label{AFFIL::GermanyUBerlin}
\item Main Astronomical Observatory of the National Academy of Sciences of Ukraine, Zabolotnoho str., 27, 03143, Kyiv, Ukraine\label{AFFIL::UkraineObsNASUkraine}
\item Space Technology Centre, AGH University of Science and Technology, Aleja Mickiewicza, 30, 30-059, Krak\'ow, Poland\label{AFFIL::PolandAGHCracowSTC}
\item Academic Computer Centre CYFRONET AGH, ul. Nawojki 11, 30-950, Krak\'ow, Poland\label{AFFIL::PolandCYFRONETAGH}
\item Institute of Astronomy, Faculty of Physics, Astronomy and Informatics, Nicolaus Copernicus University in Toru\'n, ul. Grudzi\k{a}dzka 5, 87-100 Toru\'n, Poland\label{AFFIL::PolandTorunInstituteofAstronomy}
\item Cherenkov Telescope Array Observatory gGmbH, Via Gobetti, Bologna, Italy\label{AFFIL::ItalyCTAOBologna}
\item Warsaw University of Technology, Faculty of Electronics and Information Technology, Institute of Electronic Systems, Nowowiejska 15/19, 00-665 Warsaw, Poland\label{AFFIL::PolandWUTElectronics}
\item Physics Program, Graduate School of Advanced Science and Engineering, Hiroshima University, 739-8526 Hiroshima, Japan\label{AFFIL::JapanUHiroshima}
\item School of Physics and Astronomy, Sun Yat-sen University, Zhuhai, China\label{AFFIL::ChinaUSunYatsen}
\item Department of Physical Sciences, Aoyama Gakuin University, Fuchinobe, Sagamihara, Kanagawa, 252-5258, Japan\label{AFFIL::JapanUAoyamaGakuin}
\item Division of Physics and Astronomy, Graduate School of Science, Kyoto University, Sakyo-ku, Kyoto, 606-8502, Japan\label{AFFIL::JapanUKyotoPhysicsandAstronomy}
\item Port d'Informaci\'o Cient{\'\i}fica, Edifici D, Carrer de l'Albareda, 08193 Bellaterrra (Cerdanyola del Vall\`es), Spain\label{AFFIL::SpainPIC}
\item INAF - Osservatorio Astrofisico di Torino, Strada Osservatorio 20, 10025  Pino Torinese (TO), Italy\label{AFFIL::ItalyOTorino}
\item Departamento de F{\'\i}sica, Universidad T\'ecnica Federico Santa Mar{\'\i}a, Avenida Espa\~na, 1680 Valpara{\'\i}so, Chile\label{AFFIL::ChileDepFisUTecnicaFedericoSantaMaria}
\item Faculty of Science, Ibaraki University, Mito, Ibaraki, 310-8512, Japan\label{AFFIL::JapanUIbaraki}
\end{enumerate}

%\end{document}

%first.author$^1$, 
%second.author$^2$, 
%third.author$^3$ % .... more names
%and 
%last.author$^{n}$ \\
%
%\noindent
%$^1$first.affiliation.
%$^2$second.affiliation. % .... more affiliation
%$^{m}$last.affiliation.


\begin{thebibliography}{99}
\bibitem{zwi33}F. Zwicky, Helv. Phys. Ac. {\bf 6}, 110 (1933).
\bibitem{apr18}E. Aprile {\it et al.}, Phys. Rev. Lett. {\bf 121}, 111302 (2018).
%\bibitem{tis07}P. Tisserand {\it et al.}, A\&A {\bf 469}, 387 (2007).
\bibitem{bat21}I. Batkovi{\'c} {\it et al.}, Universe {\bf 7}, 185 (2021).
\bibitem{ros04}{\L}. Roszkowski, Pramana {\bf 62}, 389 (2004).
\bibitem{ber98}L. Bergstr{\"o}m {\it et al.}, Astropart. Phys. {\bf 9}, 137 (1998).
\bibitem{cem11}J.~A.~R. Cembranos {\it et al.}, Phys. Rev. D {\bf 83}, 083507 (2011).
\bibitem{cir11}M. Cirelli {\it et al.}, JCAP {\bf 3}, 051 (2011).
\bibitem{may10}L. Mayer, Adv. Astron. {\bf 2010}, 278434 (2010).
\bibitem{eva04}N.~W. Evans {\it et al.}, Phys. Rev. D {\bf 69}, 123501 (2004).
\bibitem{str08}L.~E. Strigari {\it et al.}, ApJ {\bf 678}, 614 (2008).
\bibitem{ver18}VERITAS Coll., PoS {\bf 301}, 904 (2018).
\bibitem{mag22}MAGIC Coll., PDU {\bf 35}, 100912 (2022).
%\bibitem{cta21}CTA Cons., JCAP {\bf 01}, 057 (2021).
\bibitem{dor13}M. Doro {\it et al.}, Astropart. Phys. {\bf 43}, 189 (2013).
\bibitem{cta19}CTA Cons., {\it Science with the Cherenkov Telescope Array}, World Scientific Pub. (2019).
\bibitem{bon15}V. Bonnivard {\it et al.}, MNRAS {\bf 453}, 849 (2015).
\bibitem{hut19}M. H{\"u}tten {\it et al.}, Comp. Phys. Comm. {\bf 235}, 336 (2019).
\bibitem{ger15}A. Geringer-Sameth {\it et al.}, ApJ {\bf 801}, 74 (2015).
\bibitem{mar15}G.~D. Martinez {\it et al.}, MNRAS {\bf 451}, 2524 (2015).
\bibitem{hay16}K. Hayashi {\it et al.}, MNRAS {\bf 461}, 2914 (2016).
\bibitem{bin08}J. Binney \& S. Tremaine, {\it Galactic Dynamics: Second Edition}, Princeton Ser. (2008).
\bibitem{bon15b}V. Bonnivard {\it et al.}, MNRAS {\bf 446}, 3002 (2015).
\bibitem{ein65}J. Einasto, Tr. Astrof. Inst. Alma-Ata {\bf 5}, 87 (1965).
\bibitem{bur95}A. Burkert, ApJL {\bf 447}, L25 (1995).
\bibitem{dei17}C. Deil {\it et al.}, PoS {\bf 301}, 766 (2018).
\bibitem{ste12}G. Steigman {\it et al.}, Phys. Rev. D {\bf 86}, 023506 (2012).
\bibitem{spr08}V. Springel {\it et al.}, MNRAS {\bf 391}, 1685 (2008).
\bibitem{new17}O. Newton {\it et al.}, MNRAS {\bf 479}, 2853 (2017).
\bibitem{ive19}{\v Z}. Ivezi{\'c} {\it et al.}, ApJ {\bf 873}, 111 (2019).
\end{thebibliography}
\end{document}